\documentclass[a4paper,fleqn]{cas-dc}

\usepackage[authoryear]{natbib}
\usepackage{booktabs}
\usepackage{multirow}
\usepackage{subcaption}
\usepackage{hyperref,enumitem}
\usepackage{ulem}
\usepackage{soul}
\usepackage{float}
\usepackage{placeins}

\def\tsc#1{\csdef{#1}{\textsc{\lowercase{#1}}\xspace}}
\tsc{WGM}
\tsc{QE}
\tsc{EP}
\tsc{PMS}
\tsc{BEC}
\tsc{DE}

\begin{document}
\let\WriteBookmarks\relax
\def\floatpagepagefraction{1}
\def\textpagefraction{.001}
\shorttitle{}
\shortauthors{H. Yang et~al.}

\title [mode = title]{Text-RSIR: A Text-Guided Framework for Efficient Remote Sensing Image Transmission and Reconstruction}                      
%
%

\author[1]{Hao Yang}[orcid=0009-0008-3066-6379]
\ead{haoyang.official@gmail.com}



\affiliation[1]{organization={School of Science and Engineering, The Chinese University of Hong Kong, Shenzhen},
                city={Shenzhen},
                postcode={518172}, 
                state={Guangdong},
                country={China}}

\author[2]{Xianping Ma}[orcid=0000-0002-2180-2964]
\ead{xianpingma@link.cuhk.edu.cn}
\affiliation[2]{organization={Geosciences and Engineering, Southwest Jiaotong University},
            city={Chengdu},
            postcode={611756},
            country={China}}

\author[3]{Peifeng Ma}[orcid=0000-0002-1457-5388]
\ead{xianpingma@link.cuhk.edu.cn}
\affiliation[3]{organization={Institute of Space and Earth Information Science and the Department of Geography and Resource Management, The Chinese University of Hong Kong},
	city={Hong Kong},
	country={China}}

\author[1]{Man-On Pun}[orcid=0000-0003-3316-5381]
\cormark[1]
\ead{SimonPun@cuhk.edu.cn}
\cortext[cor1]{Corresponding author}


\begin{abstract}
	High-resolution remote sensing imagery is critical for environmental monitoring, urban mapping, and land cover analysis, but its transmission is often hindered by limited bandwidth and high communication costs. Conventional pipelines transmit full-resolution pixel data, resulting in redundant and inefficient delivery. This paper proposes a text-guided remote sensing image transmission system that replaces complete high-resolution data with low-resolution images accompanied by compact textual descriptions. An onboard text generator produces spatial and semantic summaries, reducing the transmitted data volume to approximately 2\% of the original size. For ground-based reconstruction, a text-conditioned image restoration model is introduced, which leverages cross-modal learning to recover fine spatial details and maintain semantic coherence. Experimental results on the Alsat-2B, UC Merced Land Use, and Aerial Image datasets demonstrate that the proposed framework achieves reconstruction PSNRs of 16.36 dB, 26.87 dB, and 27.41 dB, respectively, enabling efficient and information-preserving image transfer for remote sensing applications. The implementation will be made publicly available at \href{https://github.com/haoyangofficial/textrssr}{GitHub}.
\end{abstract}


\begin{highlights}
\item Proposes a text-guided remote sensing image transmission system that downlinks LR imagery with compact onboard VLM-generated captions, reducing payload size to \(\sim\)2\% of HR data.
\item Introduces Text-RSIR, a text-conditioned reconstruction network that leverages CLIP image and text embeddings and semantic guidance maps for multimodal remote sensing image reconstruction.
\item Develops an iterative text-guided feature aligning and refinement strategy with a dual-head design to jointly recover fine spatial details and preserve global semantic consistency.
\item Demonstrates consistent improvements over recent remote sensing image super-resolution baselines on Alsat-2B, UC Merced, AID, and ILSVRC2012, achieving state-of-the-art PSNR and SSIM with competitive FLOPs.
\end{highlights}

\begin{keywords}
Remote sensing image transfer \sep multimodal learning \sep text-conditioned image reconstruction
\end{keywords}

\maketitle

\section{Introduction}

Remote sensing (RS) imagery underpins a wide range of critical applications, including environmental monitoring \cite{Mahanta2024}, urban planning \cite{Huang2023}, precision agriculture \cite{Mahanta2025}, disaster response \cite{Rolla2025}, and national security. 
These tasks rely heavily on the spatial and spectral richness of high-resolution (HR) images to support reliable land-cover classification \cite{Zhang2024b}, object detection \cite{ma2024manet}, and temporal change analysis \cite{ma2024sam}. 
However, the acquisition and transmission of HR imagery are severely constrained by satellite hardware, limited onboard storage, and restricted downlink bandwidth \cite{Rong2025}. 
As the volume of RS data continues to grow, efficient transmission of high-fidelity image information has become a pressing bottleneck for global observation systems. 
Consequently, there is a growing need to rethink RS image transmission from a task- and information-centric perspective, where only the most informative content is delivered rather than exhaustive pixel-level data.

Conventional RS data pipelines prioritize the downlink of pixel-intensive HR imagery, which demands substantial communication bandwidth and incurs high operational costs \cite{Wang2025}. 
Although lossless or lossy compression can reduce data volume, these methods often compromise essential structural and semantic content required for downstream tasks. 
Moreover, environmental factors including atmospheric disturbances, cloud coverage, and acquisition delays further restrict the consistency and completeness of transmitted observations \cite{Abdelmajeed2024}, frequently resulting in partially missing or outdated information \cite{Das2025}. 
These limitations underscore the urgent need for alternative transmission strategies that preserve image fidelity while significantly reducing communication overhead for time‐critical applications. 
Therefore, reducing transmission volume while retaining task-relevant semantic and structural information remains an open and critical challenge in RS communication systems.

\begin{figure}[pos=t]
\centering

\begin{subfigure}{0.99\columnwidth}
  \centering
  \includegraphics[width=\linewidth]{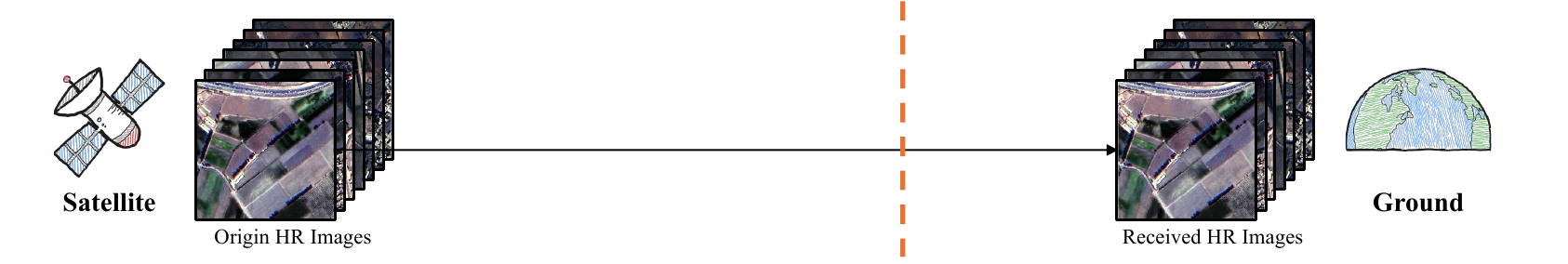}
  \caption{Conventional HR image transfer.}
  \label{fig:gap_a}
\end{subfigure}

\vspace{0.5em}

\begin{subfigure}{0.99\columnwidth}
  \centering
  \includegraphics[width=\linewidth]{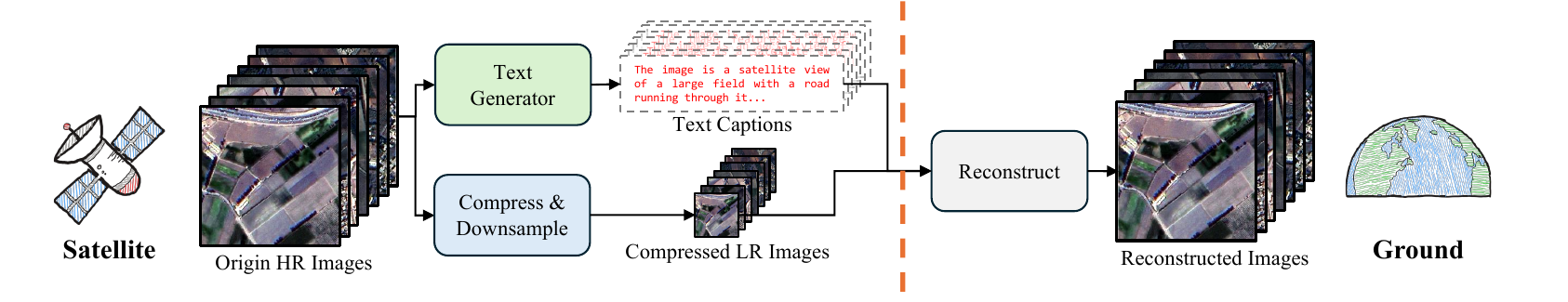}
  \caption{Proposed text-guided transfer.}
  \label{fig:gap_b}
\end{subfigure}

\caption{Comparison between conventional and text-guided RS image transfer. (a) Conventional HR image downlink. (b) Proposed transmission of LR imagery with textual descriptors for efficient and information-preserving data delivery.}
\label{fig:gap}
\end{figure}

In parallel, rapid progress in cross-modal learning and vision–language models (VLMs) \cite{clip} suggests new possibilities for improving both the informativeness and efficiency of RS data transmission \cite{Luo2025, Gandikota2024, Yao2025}. 
Rather than transferring full-resolution pixel data (as shown in Figure~\ref{fig:gap}\subref{fig:gap_a}), a more communication-efficient paradigm is to downlink lightweight yet semantically expressive representations in the form of low-resolution (LR) imagery complemented by compact textual descriptions \cite{Weng2025}, as shown in Figure~\ref{fig:gap}\subref{fig:gap_b}. 
Such captions can concisely encode high-level contextual cues that would otherwise require large HR payloads to convey, \textit{e.g.}, “a mountainous region with snow-covered peaks and sparse vegetation.” 
This paradigm shift motivates the exploration of text-guided transmission mechanisms that explicitly leverage multimodal representations to balance bandwidth efficiency and information completeness in RS data delivery.

While the proposed system enables efficient data transfer, it also introduces a new challenge: how to reconstruct or recover HR information from the transmitted LR–text pair. 
Traditionally, this problem has been addressed under the framework of remote sensing image super-resolution (RSSR), which aims to recover HR imagery from its LR counterpart by learning a mapping between LR and HR image spaces.
Existing RSSR methods can be broadly categorized into interpolation-based approaches, convolutional neural network (CNN)-based models, and more recent transformer-based architectures, all of which focus exclusively on visual signal enhancement.
Despite their effectiveness in restoring spatial details, these methods are inherently limited by the information contained in the LR image itself, making them prone to texture ambiguity, semantic inconsistency, and hallucinated artifacts when high-frequency details are severely degraded or missing.

In contrast to purely vision-based RSSR, the proposed multimodal remote sensing image reconstruction (RSIR) paradigm explicitly incorporates the textual descriptions generated onboard, providing complementary high-level semantic guidance that is unavailable in LR imagery alone.
By conditioning the reconstruction process on both visual and linguistic cues, the model can better disambiguate scene structures, preserve semantic coherence, and recover HR details in a more informed and robust manner.
To address this, this study proposes a text-conditioned remote sensing image reconstruction Network (Text-RSIR), which serves as the reception model of the transmission system, effectively bridging communication efficiency and reconstruction fidelity through multimodal alignment. 
The main contributions of this work are as follows:

\begin{enumerate}
\item This work proposes a text-guided RS Image Transmission (TGRSIT) framework, enabling low-bandwidth and information-preserving transmission via LR–text pairs;

\item This work introduces Text-RSIR, a novel text-conditioned reception model, to reconstruct HR imagery from transmitted multimodal data through iterative text-guided refinement;

\item In Text-RSIR, a dual-encoder network design is developed to simultaneously capture local visual details and maintain global semantic consistency, ensuring high-quality, semantically aligned reconstruction.
\end{enumerate}

The remainder of this paper is organized as follows: Section II reviews related work on RS image transfer, RSIR and multimodal learning. Section III details the proposed TGRSIT system and Text-RSIR framework. Section IV presents experimental setups, datasets, results, and ablation studies. Section V concludes this paper.

\section{Related work}
\subsection{Remote Sensing Image Transmission Methods}

Remote sensing image transmission has traditionally been dominated by pixel-centric downlink pipelines, where bitrate reduction is achieved primarily through onboard compression while maintaining the original spatial resolution and radiometric consistency of the acquired imagery. 
In operational satellite systems, transform-based codecs such as JPEG2000 and standardized solutions such as the CCSDS Image Data Compression recommendation are widely adopted because they provide predictable rate--distortion trade-offs under strict onboard memory, power, and bandwidth constraints \cite{taubman2002jpeg2000,ccsds122b2}. 
However, these approaches fundamentally preserve the full-resolution pixel grid and focus exclusively on removing statistical redundancy, resulting in limited compression efficiency for large-scale RS scenes that contain extensive spatial repetition and semantically redundant structures. 
Moreover, prior studies have shown that lossy compression can significantly influence downstream interpretation tasks, such as land-cover classification and object recognition, even when conventional fidelity metrics remain high \cite{zabala2012impact}. 
As a result, traditional compression methods neither explicitly account for task relevance nor selectively prioritize semantically informative content, leading to incomplete exploitation of bandwidth reduction opportunities and suboptimal task-oriented performance.

Recent advances in learned image compression have further improved rate-distortion efficiency by replacing hand-crafted transforms with end-to-end optimized neural latent representations and more expressive entropy models \cite{balle2018hyperprior,minnen2018joint}. 
While these approaches outperform classical codecs in many scenarios, they remain fundamentally pixel-driven and typically operate at the native spatial resolution of the input image. 
Consequently, they inherit similar limitations in terms of semantic redundancy and task agnosticism, as the transmitted representation is still optimized primarily for image reconstruction rather than downstream decision-making. 
In parallel, task-oriented and semantic communication paradigms have emerged, advocating the transmission of task-relevant features or representations instead of raw imagery, and demonstrating clear advantages under bandwidth-limited conditions \cite{kang2021task}. 
Nevertheless, most existing task-oriented RS transmission methods rely on lightweight feature encoders tailored to specific downstream tasks, limiting their generality and flexibility across diverse RS applications.
At the same time, rapid progress in edge AI hardware and model optimization has made the deployment of VLMs increasingly feasible in resource-constrained environments. 
Recent industrial reports and demonstrations indicate that models on the order of several billion parameters can be executed on embedded GPU platforms through a combination of mixed-precision inference, parameter-efficient architectures, and software-hardware co-design \cite{nvidia_orin}. 
In particular, NVIDIA’s Jetson-class systems and space-qualified GPU roadmaps highlight a clear trend toward onboard execution of foundation models for perception and reasoning tasks, including multimodal understanding \cite{llm_edge}. 
These developments suggest that integrating VLMs with satellite payloads is becoming technically viable, enabling semantic abstraction and reasoning to be performed directly at the data source rather than solely on the ground.

Building upon these trends, our approach combines the complementary strengths of conventional compression and semantic-aware transmission. 
First, instead of relying solely on lossy compression at the original resolution, we introduce explicit spatial downsampling as an additional and orthogonal means of reducing data volume, substantially lowering the payload size before compression while accepting controlled loss of fine-grained detail. 
Second, to compensate for the information discarded during downsampling, an onboard VLM is employed to generate compact textual descriptions that explicitly encode high-level semantic and structural cues. 
By transmitting LR imagery together with lightweight language descriptions, the proposed framework departs from purely pixel-centric delivery and enables the receiver to reconstruct high-resolution content in a semantically informed manner. 
This hybrid design bridges traditional rate-distortion–oriented compression and emerging task-oriented communication, offering a practical and scalable solution for bandwidth-limited remote sensing systems.

\subsection{Remote Sensing Image Reconstruction Methods}

With increasing constraints on bandwidth, onboard storage, and sensor capabilities, computational reconstruction has become essential for maximizing the utility of transmitted RS data \cite{He2025}. 
RSIR refers to a broad class of methods that recover high-quality RS imagery from degraded, incomplete, or compressed observations, encompassing tasks such as super-resolution (SR), denoising, deblurring, inpainting, and compressed reconstruction \cite{Wang2022b, Yang2025}. 
These methods aim to restore spatial, spectral, and semantic fidelity under realistic acquisition and transmission constraints.

Among RSIR tasks, RSSR has been the most extensively studied, focusing on reconstructing HR images from LR inputs to recover fine spatial details critical for downstream analysis \cite{He2025}. 
Early RSIR and RSSR approaches relied on signal-processing and statistical models. 
Interpolation-based methods, such as bicubic interpolation \cite{bicubic}, provide simple baselines but produce overly smooth results due to missing high-frequency information. 
Sparse representation and dictionary learning methods \cite{Yang2010} improve detail recovery by exploiting learned LR–HR patch correspondences, yet remain limited in modeling the complex spatial heterogeneity of RS scenes.

Recent advances in deep learning have significantly improved RSIR performance by learning non-linear reconstruction mappings directly from data \cite{He2025, Wang2022b}. 
CNN-based models, including SRCNN \cite{srcnn} and EDSR \cite{edsr}, enabled early end-to-end RSSR and were later extended to other reconstruction tasks such as denoising and deblurring \cite{Zhang2024, Li2024}. 
GAN-based approaches, such as SRGAN \cite{srgan}, NDSRGAN \cite{ndsrgan}, Dual RSS-GAN \cite{dualrssgan}, and CDGAN \cite{cdgan}, further enhance perceptual quality through adversarial learning. 
More recently, transformer-based architectures introduce global self-attention to capture long-range dependencies in large-scale RS scenes \cite{Liu2025}, while diffusion-based models, including SRDiff \cite{srdiff} and DDRM \cite{ddrm}, formulate RSIR as a probabilistic refinement process. 
State-space and Mamba-based designs, such as FMSR \cite{fmsr} and EMAN \cite{eman}, further improve long-range modeling efficiency. 
These developments highlight the increasing architectural diversity of modern RSIR methods \cite{Wang2022b}.

Despite their success, most existing RSIR and RSSR methods remain fundamentally unimodal, relying exclusively on visual signals and often assuming unrestricted access to HR data during training and inference. 
Such assumptions are misaligned with practical RS systems constrained by limited bandwidth, onboard processing budgets, and incomplete data availability \cite{Yang2025}. 
Moreover, purely pixel-driven reconstruction lacks high-level semantic awareness, which can lead to ambiguity and semantic inconsistency under severe degradation.

To overcome these limitations, this work proposes Text-RSIR, a multimodal RSIR framework that incorporates textual semantics as auxiliary supervisory information. 
In the proposed system, an onboard VLM generates semantic captions from HR observations, which are encoded using CLIP embeddings \cite{clip} and transmitted alongside LR imagery. 
At the receiver, CNN-based visual features are guided by text-derived semantic representations to enhance structural consistency and semantic alignment during reconstruction \cite{segsr}. 
By leveraging multimodal priors, Text-RSIR extends conventional RSSR toward a more general, context-aware RSIR paradigm, enabling semantically enriched and bandwidth-efficient reconstruction under realistic RS transmission constraints.

\subsection{Multimodal Learning in Remote Sensing Image Reconstrction}

Multimodal Learning in RSIR refers to approaches that leverage complementary data types, such as Synthetic Aperture Radar (SAR), optical imagery, hyperspectral and multispectral images, or even textual/contextual information, to enhance the spatial and/or spectral fidelity of reconstructed HR images.

Multimodal learning-based RSIR methods can broadly be divided into cross-modal fusion methods and joint embedding or contrastive learning methods. 
Cross-modal fusion approaches seek to integrate complementary information from multiple sensors to enhance reconstruction quality. 
For example, the Multi-Resolution Collaborative Fusion of SAR, Multispectral and Hyperspectral Images method combines SAR, multispectral, and hyperspectral imagery through deep fusion strategies to preserve spatial, spectral, and geometrical fidelity in reconstructed outputs, demonstrating effective cross-modal integration in RSSR contexts \cite{Yuan2022}. 
Another example involves combining LR hyperspectral with HR multispectral data through transformer-based fusion or retractable spatial–spectral modeling, yielding better spectral and spatial details \cite{Hu2021}. 
In contrast, joint embedding and contrastive learning methods focus on learning a shared latent representation across modalities. 
One representative example is the multimodal contrastive learning framework for RS tasks, which employs a dual-encoder architecture pre-trained on approximately one million paired Sentinel-1 (radar) and Sentinel-2 (optical) image pairs \cite{Jain2022}. 
This framework aligns features across modalities via contrastive learning and shows improved performance in downstream tasks like segmentation and land-cover mapping. 
Another compelling approach is the Contrastive Radar‑Optical Masked Autoencoders, which fuses cross‑modal contrastive learning with masked autoencoding to learn rich multimodal representations from spatially aligned SAR and multispectral optical imagery \cite{Fuller2023}. 
Despite the promise of these fusion and embedding‑based RSIR approaches, they tend to rely heavily on additional visual modalities (such as SAR, hyperspectral, or multispectral data), while largely neglecting the potential of textual information, including captions and semantic descriptions, that could provide high‑level guidance for more semantically aware super‑resolution.

To address this gap, this study introduces Text-RSIR, which synergizes image data with auxiliary textual and semantic cues by first generating captions for HR images, downsampling them, and encoding the LR images via CLIP embeddings. 
By explicitly incorporating these auxiliary modalities, Text-RSIR remedies the underutilization of language in multimodal RSIR and enables more semantically faithful reconstruction. 
While no prior RSIR method explicitly employs text conditioning, related work in natural image SR demonstrates the viability of CLIP-based text guidance, as in the recent Text-Guided Explorable Image Super-Resolution framework \cite{Gandikota2024}. 
Text-RSIR thus represents a novel extension of multimodal learning to RSIR, integrating visual and linguistic modalities for the first time.

\section{Proposed Method}

\subsection{Overview}

\begin{figure*}[pos=t]
\centering
\includegraphics[width=\textwidth]{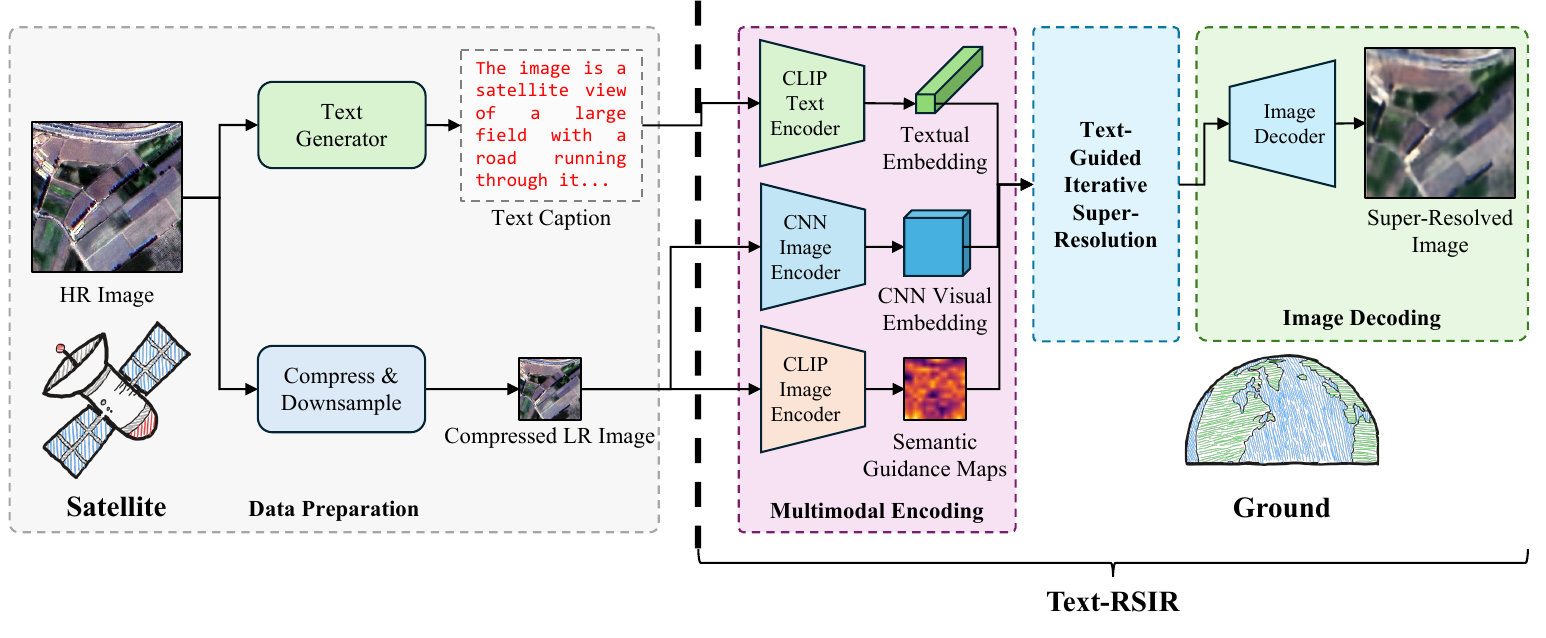}
\caption{Overall architecture of the proposed TGRSIT. The system transmits lightweight LR–text pairs from satellite to ground, where Text-RSIR reconstructs HR imagery guided by semantic information.}
\label{fig:concept}
\end{figure*}

This study proposes TGRSIT, in which Text-RSIR serves as the intelligent decoder, as illustrated in Figure \ref{fig:concept}.
As shown in the figure, the proposed transmission system consists of four main stages: Data Preparation, Multimodal Encoding, Text-Guided Iterative Super-Resolution (TGISR), and Image Decoding. 
Among these stages, Data Preparation is performed on the satellite, while Text-RSIR (corresponding to the latter three stages) is implemented on the ground.

During data preparation, the captured HR image is first processed by the Text Generator (a lightweight VLM) to produce a text caption (TC).
The image is then downsampled and compressed.
The resulting compressed low-resolution (CLR) image, along with the TC, is transmitted to the ground.

\begin{figure*}[pos=t]
\centering
\includegraphics[width=\textwidth]{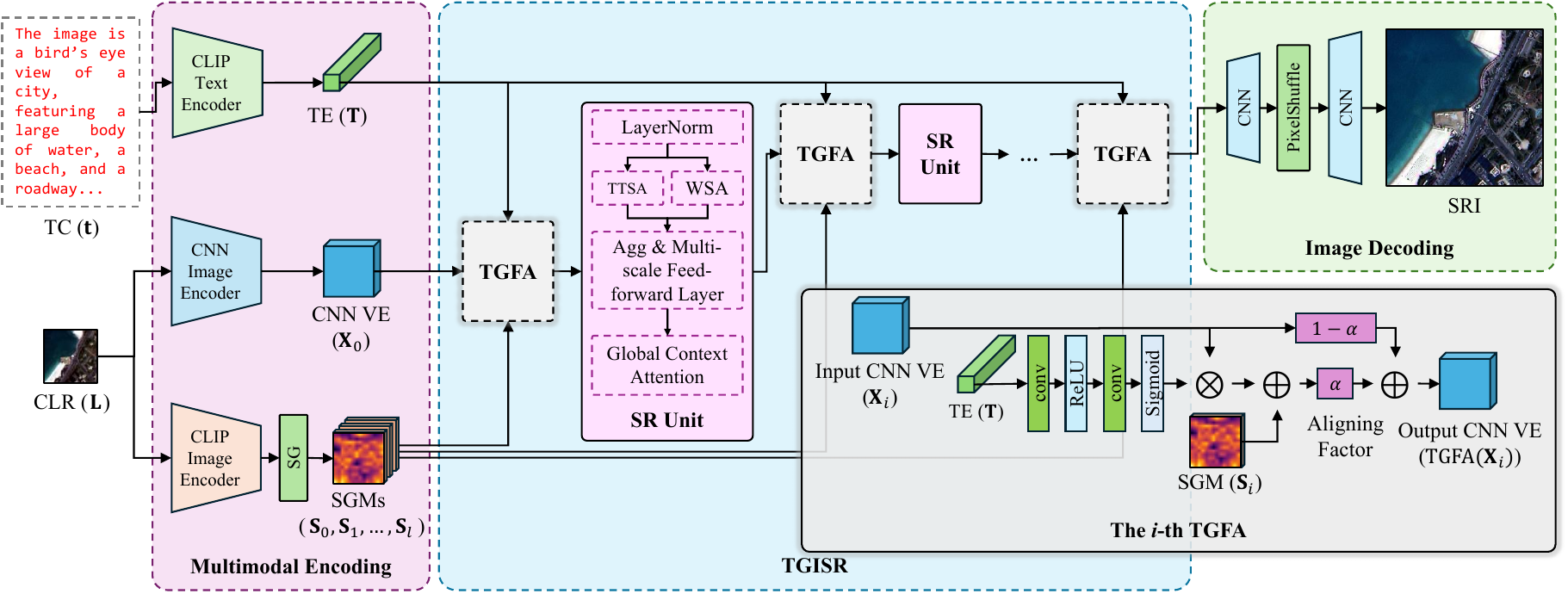}
\caption{Overview of the proposed Text-RSIR. TTSA denotes Top-$k$ Token Selective Attention, and WSA represents Window-based Self-Attention within the TTST architecture.}
\label{fig:archi}
\end{figure*}

Text-RSIR is designed to enhance RSIR performance through the iterative utilization of the TC and the CLIP embeddings. 
Figure \ref{fig:archi} presents further details of the Text-RSIR architecture.
During Multimodal Encoding, the LR image is processed by two separate encoders (a CNN Image Encoder and a CLIP Image Encoder) to produce a CNN Visual Embedding (CNN VE) and CLIP Visual Embeddings (CLIP VEs), respectively.
The CLIP VEs are subsequently utilized to generate semantic guidance maps (SGMs).
Simultaneously, the TC is processed through a CLIP Text Encoder to generate a Textual Embedding (TE). 
In TGISR, the CNN VE is iteratively aligned and refined. 
In each iteration, all embeddings are input into a Text-Guided Feature Aligning (TGFA) step, followed by processing through an SR Unit. 
After the final iteration, the refined CNN VE goes through one last TGFA. 
Finally, in the Image Decoding step, the processed CNN VE is decoded by a CNN decoder, upsampled, and further decoded by an additional CNN.

\subsection{Data Preparation}

In the Data Preparation step, the HR image captured by the satellite is first downsampled and compressed to produce a CLR image, which serves as the input for image reconstruction. 
At the same time, the HR image is used to generate a descriptive TC that summarizes its semantic content, such as terrain type, objects, or structural patterns. 
This TC is created by a VLM running onboard the satellite, allowing semantic information to be extracted before data transmission.

Since transmitting full HR images requires considerable bandwidth and storage, only the CLR images and the lightweight TCs are sent to the Earth. 
This design not only reduces communication cost but also ensures that semantic details are preserved through the captions, compensating for information lost during downsampling. 
As a result, the prepared data provide both spatial structure from the CLR images and contextual knowledge from the captions, forming the foundation for multimodal SR.

\subsection{Multimodal Encoding}

In this process, the CLR image is processed by both a CNN Image Encoder and a CLIP Image Encoder to generate the CNN VE and SGMs, respectively. At the same time, the TC is processed through a CLIP Text Encoder to produce the TE.

As discussed, RS images often contain repetitive patterns and similar scene types, making it difficult for the CLIP Image Encoder alone to effectively distinguish between them. To address this, a dual image encoder architecture is adopted to combine the semantic understanding capability of the CLIP Image Encoder with the local feature extraction strength of the CNN. In this architecture, the CNN VE serves as the primary feature, while the SGMs generated from CLIP VEs provide supplementary information to guide feature adjustment and aligning.

The image encoding process is described as follows. Let $\textbf{L} \in \mathbb{R}^{C \times H \times W}$ represent the CLR image received from the satellite. The extraction of the CNN VE using the CNN Image Encoder is defined by:
\begin{equation}
\textbf{X}_0 = \text{IE}_\text{CNN}(\textbf{L}),
\end{equation}
where $\text{IE}_\text{CNN}$ denotes the CNN Image Encoder, and $\textbf{X}_0$ represents the resulting CNN VE.

To provide CLIP-based semantic guidance, the first token and the remaining tokens from the CLIP Image Encoder output are separated. 
We adopt the Semantic Guiding (SG) process from SeG-SR \cite{segsr} to generate an SGM for each TGFA step in the TGISR process. 
The SG process is designed to modulate local feature enhancement in the SR pipeline using semantic priors from CLIP. For each step $i \in [0, l]$, the SG process computes an SGM $\mathbf{S}_i \in \mathbb{R}^{B \times H\times W}$ for each of the $l$ SR Units, providing explicit spatially-aware guidance aligned with the target semantics.

Let $\mathbf{v}_0 \in \mathbb{R}^{B \times C}$ denote the global semantic embedding (\textit{i.e.}, the [CLS] token) and $\mathbf{v}_\text{rm} \in \mathbb{R}^{B \times (H \times W) \times C}$ the local semantic embeddings extracted from the CLIP Image Encoder. The SG process for the $i$-th step is defined as:
\begin{equation}
\text{SG}_i(\mathbf{v}_0, \mathbf{v}_\text{rm}) = \text{CS}(\Theta(\mathbf{e}_i), \Theta(\mathbf{v}_\text{rm})),
\end{equation}
where $\text{CS}(\cdot, \cdot)$ denotes the cosine similarity, and $\Theta(\cdot)$ is the $\ell_2$-normalization function applied along the feature dimension. The latent embeddings $\mathbf{e}_i \in \mathbb{R}^{B \times (l+1) \times C}$ used for computing the SGMs are derived through the following stages \cite{metanet}:
\begin{equation}
\begin{aligned}
\left\{
\begin{aligned}
\mathbf{p}_\text{a} &= \text{MetaNet}(\mathbf{v}_0, \mathbf{p}_0), \\
[\mathbf{p}_\text{a}^\ast; \mathbf{p}_\text{b}^\ast] &= \text{MHSA}([\mathbf{p}_\text{a};\mathbf{p}_\text{b}]),\\
[\mathbf{e}_i] &= \text{MLP}_\textbf{e}(\text{MLP}_\text{a}(\mathbf{p}_\text{a}^\ast) \odot \sigma(\text{MLP}_\text{b}(\mathbf{p}_\text{b}^\ast))).
\end{aligned}
\right.
\end{aligned}
\end{equation}
Here, $\mathbf{p}_0 \in \mathbb{R}^{1 \times C}$ is a learnable global prior descriptor, and $\mathbf{p}_\text{b} \in \mathbb{R}^{(l+1) \times C}$ are learnable local unit descriptors, shared across all steps, and $\odot$ represents element-wise multiplication. $\text{MetaNet}$ is an architecture to extract semantic correspondence between $\mathbf{v}_0$ and $\mathbf{p}_0$. The Multi-Head Self-Attention (MHSA) block refines the concatenated descriptors $[\mathbf{p}_\text{a}; \mathbf{p}_\text{b}]$, enabling context exchange between global and unit tokens \cite{mhsa}. The gating mechanism modulates the unit embeddings via element-wise product between the transformed global vector and a gated local vector. The resulting fused descriptors $\mathbf{e}_i$ are compared with the normalized local features $\mathbf{v}_\text{rm}$ via cosine similarity, and $i\in[0,l]$.

\begin{figure}[pos=t]
\centering
\includegraphics[width=\columnwidth]{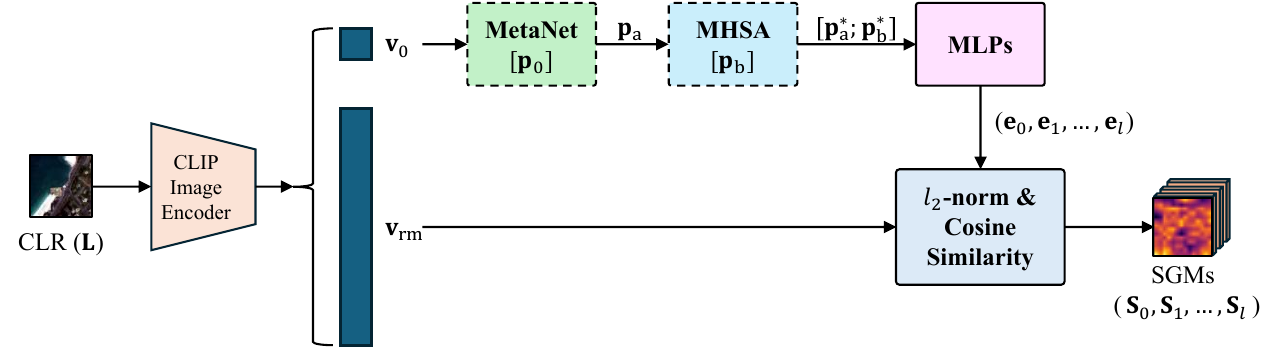}
\caption{Overview of the SG process. The MetaNet and MHSA blocks are shown in dashed boxes, indicating that they contain learnable parameters, which are specified in square brackets following their names.}
\label{fig:sg}
\end{figure}

With the help of the SG process, the extraction of CLIP VEs and the computation of the SGM for the $i$-th TGFA step are defined as follows:
\begin{equation}
\textbf{S}_i = \text{SG}_i(\textbf{v}_0, \textbf{v}_\text{rm}),
\end{equation}
where
\begin{equation*}
\textbf{v}_0, \textbf{v}_\text{rm} = \text{IE}_\text{CLIP}(\textbf{L}).
\end{equation*}
Here, $\textbf{S}_i$ and $\text{SG}_i$ denote the SGM and the SG process for the $i$-th TGFA step, respectively. $\text{IE}_\text{CLIP}$ denotes the CLIP Image Encoder, $\textbf{v}_0$ represents the first token of the resulting CLIP VEs, and $\textbf{v}_\text{rm}$ denotes the remaining tokens. Let the number of SR Units be $l$. Then, the number of TGFA steps is $l+1$, with the index $i \in [0, l]$.
The process is illustrated in Figure \ref{fig:sg}.

The output of the SG process is an SGM $\mathbf{S}_i \in \mathbb{R}^{B \times H \times W}$, where $H \times W$ corresponds to the resolution of the CLR image. This map serves as a soft spatial prior for the $i$-th TGFA step, guiding the model to enhance semantically meaningful regions in accordance with the text prompt. This design allows each SR Unit to adaptively incorporate global semantic context and local alignment cues, thereby improving spatial correspondence and preserving high-frequency details during the SR reconstruction process.

The TE is generated using the CLIP Text Encoder as follows:
\begin{equation}
\textbf{T} = \text{TE}_\text{CLIP}(\textbf{t}),
\end{equation}
where $\textbf{t}$ denotes the caption of the HR image generated by the VLM, $\text{TE}_\text{CLIP}$ denotes Text Encoder, and $\textbf{T}$ represents the resulting TE.

\subsection{Text-Guided Iterative Super-Resolution}

In this step, the CNN VE is iteratively aligned and refined. To maintain the supervisory influence of the SGMs and TE throughout the entire SR process, each iteration begins by using these embeddings to align the CNN VE. The aligned CNN VE is then processed through an SR Unit. After the final iteration, the refined CNN VE undergoes one final TGFA step. Let the CNN VE input for the $i$-th iteration be $\textbf{X}_i$, where $i \in [0, l)$. Then, the $i$-th iteration is defined as:
\begin{equation}
\textbf{X}_{i+1} = \text{SR}(\text{TGFA}(\textbf{X}_i)),
\end{equation}
where $\text{TGFA}$ denotes the TGFA step, and $\text{SR}$ denotes the SR Unit.

TGFA is designed to align and refine the CNN VE using both the SGMs and the TE. In this process, textual information provides channel-wise attention, enabling the model to identify and emphasize the most relevant feature channels for the SR task. Following this channel-wise refinement, the SGMs are applied to perform pixel-wise refinement, offering direct high-frequency guidance to enhance spatial detail in the reconstructed image.

Let $\textbf{T} \in \mathbb{R}^{d}$ be the TE, $\textbf{S}_i$ the SGM for the $i$-th iteration, and $d$ the dimension of CLIP’s hidden state. The TGFA process is defined as:
\begin{equation}
\text{TGFA}(\textbf{X}_i) = \alpha \cdot \text{Align}(\textbf{X}_i) + (1 - \alpha) \cdot \textbf{X}_i,
\end{equation}
where
\begin{equation*}
\text{Align}(\textbf{X}_i) = \textbf{X}_i \odot \text{TG}(\textbf{T}) + \textbf{S}_i.
\end{equation*}
Here, $\alpha$ is a learnable aligning factor that controls the proportion of alignment, and $\text{Align}(\textbf{X}_i)$ denotes the aligned visual embeddings. The function $\text{TG}$ is a lightweight MLP composed of a convolution layer, a ReLU activation, another convolution layer, and a Sigmoid activation, projecting the TE into the channel space.

The complete process of TGFA is depicted in the shaded gray area of Figure \ref{fig:archi}. For the SR Unit, we adopt the Residual Token Selective Group (RTSG) from the Top-$k$ Token Selective Transformer (TTST) \cite{ttst} architecture.

\subsection{Image Decoding}

In the Image Decoding stage, Text-RSIR transforms the final text-aligned visual representation into a super-resolved image (SRI) using a lightweight CNN-based upsampling head. 
After the last TGFA step, the resulting CNN VE is first refined by a convolutional module that aggregates local spatial structures and suppresses residual artifacts from the iterative reconstruction process. 
The refined feature is then integrated into a residual decoding pathway, where shallow image features are fused with deep representations to preserve low-frequency information and enhance reconstruction stability. 
This residual formulation allows the decoder to maintain radiometric consistency while directing deeper features toward recovering high-frequency details.

For spatial resolution recovery, the decoder employs a learnable pixelshuffle-style upsampling strategy, which increases resolution by rearranging channel information into the spatial domain. 
This approach enables content-adaptive upscaling and more accurate restoration of edges and textures compared to fixed interpolation. 
A final CNN subsequently maps the upsampled features to the image space, producing the super-resolved output. 
Through sequential feature refinement, residual fusion, and learnable upsampling, the decoder effectively converts multimodally enhanced embeddings into spatially detailed and semantically coherent HR reconstructions.

\section{Experiments}
\subsection{Implementation details}

This study is implemented using PyTorch \cite{pytorch}, and all experiments are conducted on a system equipped with an Intel Core i9-10900X CPU, 128 GiB of RAM, and a single NVIDIA GeForce RTX 4090 GPU. 
The operating system is Ubuntu 22.04 (Jammy) with CUDA version 12.4. 
The downscale factor used to generate the CLR images is set to 4.
The Text-RSIR framework, along with all baseline models used for comparison, adopts L1 Loss \cite{l1loss} as the loss function and employs the AdamW optimizer \cite{adam} with an initial learning rate of $10^{-4}$. 
The learning rate is reduced by half at the midpoint of the training process. 
In this study, the training epochs were set to a moderate number to simulate a computation-constrained scenario. 
All models are trained from scratch, except for the CLIP components. 
In TGRSIT, to minimize the computational burden on the satellite while maintaining high-quality text generation, we adopt LLaVa-1.5 7B as the text generator \cite{llava}.
For Text-RSIR, the CLIP Image Encoder and Textual Encoder used in the Multimodal Encoding step are pretrained and kept frozen during training. 
The initial aligning factor $\alpha$ in the TGFA step is set to 0.5. 
The number of SR Units, $l$, is set to 6. 
The RTSG SR Unit is initialized following the configuration provided in the original TTST paper \cite{ttst}. 
Model performance is assessed using Peak Signal-to-Noise Ratio (PSNR, calculated on the RGB channels) \cite{psnr} and Structural Similarity Index Measure (SSIM) \cite{ssim}. 

\subsection{Dataset Description}

Benchmark evaluations are conducted on three public RSSR datasets: Alsat-2B dataset \cite{Alsat}, UC Merced Land Use Dataset \cite{ucmerced}, and Aerial Image Dataset (AID) \cite{aid}; as well as on a common object dataset, the ImageNet Large Scale Visual Recognition Challenge 2012 (ILSVRC2012) dataset \cite{ILSVRC15}. 

The Alsat-2B dataset \cite{Alsat} exhibits significant color style discrepancies between HR and LR images, posing additional challenges for the RSSR task.
The UC Merced Land Use Dataset \cite{ucmerced} comprises 21 categories, each containing 100 HR images.
In this study, 75 images per category are randomly selected for training, with the remaining 25 used for testing.
The AID \cite{aid} includes 30 scene categories, each with several hundred images and an original resolution of $600\times 600$.
Here, 80\% of the images from each category are randomly chosen for training, and the remaining 20\% for testing.
Lastly, the ILSVRC2012 dataset \cite{ILSVRC15} is a widely used ImageNet subset containing 1,000 object categories.
For training, 20 images per category are randomly selected, and for testing, 1,000 images are randomly drawn from the original ILSVRC2012 test set.
For the UC Merced Land Use Dataset, the HR images are provided in lossless TIF format, while the CLR images are obtained by downsampling the HR images and converting them to JPG format.
For the other three datasets, the CLR images are generated by downsampling the HR images, with both HR and CLR images stored in JPG format.
Comprehensive dataset statistics are provided in Table \ref{tab:datasets}.

\begin{table*}[pos=t]
  \centering
  \caption{Detailed information on all four datasets. “Format” refers to the file formats used for the LR and HR images. “Res.” stands for resolution, and “Sam.” denotes the number of samples. “UC Merced” refers to the UC Merced Land Use Dataset. “Image Res.” indicates the exact resolution of the HR images used in the experiments. For AID, center-cropping is applied. For ILSVRC2012, images are first resized so that the shorter edge is 256 pixels, followed by center-cropping.}
  \resizebox{\textwidth}{!}{%
  \begin{tabular}{ccccccccccc}
    \toprule
    \textbf{Dataset} & \textbf{Format} & \textbf{Spatial Res.} & \textbf{Image Res.} & \textbf{Training Sam.} & \textbf{Testing Sam.}\\
    \midrule
    Alsat-2B & JPG\textrightarrow JPG & 2.5m & $256\times 256$ & 2,182 & 577 \\
    UC Merced & JPG\textrightarrow TIF & 0.3m & $256\times 256$ & 1,575 & 525 \\
    AID & JPG\textrightarrow JPG & 0.5m & $576\times 576$ & 8,000 & 2,000 \\
    ILSVRC2012 & JPG\textrightarrow JPG & - & $256\times 256$ & 20,000 & 1,000 \\
    \bottomrule
  \end{tabular}
  }
  \label{tab:datasets}
\end{table*}

\subsection{Performance Comparison}

To evaluate the compression performance of the proposed TGRSIT system, we compare the downlink file sizes of various transmission methods, including the proposed approach, using the UC Merced Land Use Dataset.
We also implement a baseline RSIR method to reconstruct the images and report the PSNR and SSIM values of the reconstructed images relative to the lossless HR images.
For the TGRSIT, the corresponding RSIR baseline is Text-RSIR, while for the other downlink methods, the baseline used is TTST \cite{ttst}.
JPEG is employed as the compression method in TGRSIT.
The quantitative results are presented in Table \ref{tab:system}.
From the table, we observe that downsampling reduces the downlink file size to approximately 6\% of the original, while JPEG compression further reduces it to about 1.8\%. 
However, this increased compression comes at the cost of reconstruction accuracy, resulting in a PSNR drop of around 0.6 dB and an SSIM decrease of approximately 0.03.
Introducing text guidance mitigates this performance degradation, yielding a PSNR gain of 0.04 dB with only a 0.7 MB increase in file size over the 418.2 MB full-size dataset.
In summary, the proposed TGRSIT leverages text captions to preserve the original image content, providing an effective and efficient solution for satellite image transmission.

\begin{table*}[pos=t]
  \centering
  \caption{Quantitative results of the experiments on different downlink methods. “File Size” indicates the total file size that must be transmitted during downlink. “Reconstruct” is abbreviated as “Rec.”, and the reconstructed PSNR and SSIM values are computed using the corresponding baseline RSIR methods.}
  \resizebox{\textwidth}{!}{%
  \begin{tabular}{ccccccccccc}
    \toprule
    \textbf{Downlink Method} & \textbf{Downlink File Format} & \textbf{File Size} (MB) & \textbf{Rec. PSNR} (dB) & \textbf{Rec. SSIM}\\
    \midrule
    HR & TIF & 418.2 & $\infty$ & 1.0000 \\
    Downsample & TIF & 24.9 & 27.4235 & 0.7738 \\
    Downsample + Compression & JPG & 7.5 & 26.8258 & 0.7440 \\
    TGRSIT & JPG + TXT & 8.2 & 26.8675 & 0.7475 \\
    \bottomrule
  \end{tabular}
  }
  \label{tab:system}
\end{table*}

To the best of our knowledge, there is no existing RSIR method that leverages text information. 
Therefore, we compare the performance of Text-RSIR with state-of-the-art single-modal RSIR methods, including FAT \cite{fat}, FMSR \cite{fmsr}, GCRDN \cite{gcrdn}, TSFNet \cite{tsfnet}, and TTST \cite{ttst}, as all of them are compatible with image reconstruction in the transmission system.
To simulate a computation-restricted scenario, all models are trained for a moderate number of epochs (specifically, 50, 100, 20, and 20 epochs) on the four datasets, respectively.
The batch size for Text-RSIR is set to 1 on the AID and 4 for the other three datasets.

As shown in Tables \ref{tab:alsat_result}, \ref{tab:ucmerced_result}, \ref{tab:aid_result} and \ref{tab:imagenet_result}, Text-RSIR achieves the highest PSNR and SSIM scores across all three datasets. 
The learnable parameter counts and FLOPs for each model are provided in Table \ref{tab:alsat_result}. 
These values are computed using Python’s thop library, with FLOPs calculated for a $64 \times 64$ RGB input image. 
Compared with the TTST model, Text-RSIR improves PSNR by 0.0878 dB and SSIM by 0.0160 on the Alsat-2B dataset, while increasing FLOPs by only 0.02 G. 
This highlights Text-RSIR’s high computational efficiency. 
Due to the domain shift between LR and HR images in the Alsat-2B dataset, FAT and GCRDN exhibit a substantial drop in performance (PSNR $\leq16$ dB), indicating their limited ability to handle domain shift. 
Moreover, in the UC Merced Land Use Dataset and AID, which include a greater variety of image categories, the benefits of text guidance are more pronounced. 
In the UC Merced Land Use Dataset, in which we implement compression on LR images, Text-RSIR achieves a PSNR of 27.4904 dB and an SSIM of 0.7775, surpassing the second-best results by 0.669 dB and 0.0037, respectively. In AID, which comprises 30 image categories, Text-RSIR attains a PSNR of 27.4089 dB and an SSIM of 0.7116, exceeding the second-best results by 0.0742 dB and 0.0024, respectively. 
These quantitative findings underscore the effectiveness of incorporating text guidance.

\begin{table*}[pos=t]
  \centering
  \caption{Quantitative Results of the experiment on the Alsat-2B dataset. “Param.” indicates the learnable parameter count of the model. The best result is highlighted in \textbf{bold}, while the second-best result is \underline{underlined}. The unit for PSNR is decibels (dB).}
  \resizebox{\textwidth}{!}{%
  \begin{tabular}{ccccccccccc}
    \toprule
    \multirow{3}{*}{\textbf{Model}} & \multirow{3}{*}{\textbf{Param.} (M)} & \multirow{3}{*}{\textbf{FLOPs} (G)} & \multicolumn{2}{c}{\textbf{Average}} & \multicolumn{2}{c}{\textbf{Agriculture}} & \multicolumn{2}{c}{\textbf{Special}} & \multicolumn{2}{c}{\textbf{Urban}}\\
    \cmidrule(lr){4-5} \cmidrule(lr){6-7} \cmidrule(lr){8-9} \cmidrule(lr){10-11}
    & & & \textbf{PSNR} & \textbf{SSIM} & \textbf{PSNR} & \textbf{SSIM} & \textbf{PSNR} & \textbf{SSIM} & \textbf{PSNR} & \textbf{SSIM}\\
    \midrule
    FAT & 10.19 & 43.86 & 16.0099 & \textbf{0.3707} & 17.5152 & \textbf{0.4409} & 16.6609 & \underline{0.3946} & 14.8890 & \underline{0.3261}\\
    FMSR & 7.05 & 30.71 & 16.2615 & 0.3539 & 17.6293 & 0.4184 & 16.8002 & 0.3768 & 15.3054 & 0.3117\\
    GCRDN & 31.83 & 443.80 & 15.8084 & 0.2875 & 17.1307 & 0.3931 & 16.3803 & 0.3188 & 14.8237 & 0.2258\\
    TSFNet & 3.13 & 111.17 & \textbf{16.4214} & \underline{0.3704} & \underline{17.7688} & 0.4376 & \textbf{17.0326} & \textbf{0.3951} & \underline{15.3846} & 0.3255 \\
    TTST & 18.30 & 76.84 & 16.2754 & 0.3542 & 17.6870 & 0.4220 & 16.8012 & 0.3764 & 15.3243 & 0.3121\\
    Text-RSIR & 22.72 & 76.86 & \underline{16.3632} & 0.3702 & \textbf{18.0134} & \underline{0.4385} & \underline{16.8149} & 0.3868 & \textbf{15.4435} & \textbf{0.3346}\\
    \bottomrule
  \end{tabular}
  }
  \label{tab:alsat_result}
\end{table*}

\begin{table*}[pos=t]
  \centering
  \caption{Quantitative Results of the experiment on the UC Merced Land Use Dataset. Note that the LR images are compressed using JPEG. The unit for PSNR is decibels (dB).}
  \resizebox{\textwidth}{!}{%
  \begin{tabular}{ccccccccccccc}
    \toprule
    \multirow{3}{*}{\textbf{Class}} 
    & \multicolumn{2}{c}{\textbf{FAT}} 
    & \multicolumn{2}{c}{\textbf{FMSR}}
    & \multicolumn{2}{c}{\textbf{GCRDN}}
    & \multicolumn{2}{c}{\textbf{TSFNet}}
    & \multicolumn{2}{c}{\textbf{TTST}}
    & \multicolumn{2}{c}{\textbf{Text-RSIR}} \\
    \cmidrule(lr){2-3} \cmidrule(lr){4-5} \cmidrule(lr){6-7} \cmidrule(lr){8-9} \cmidrule(lr){10-11} \cmidrule(lr){12-13}
    & \textbf{PSNR} & \textbf{SSIM} 
    & \textbf{PSNR} & \textbf{SSIM} 
    & \textbf{PSNR} & \textbf{SSIM} 
    & \textbf{PSNR} & \textbf{SSIM} 
    & \textbf{PSNR} & \textbf{SSIM} 
    & \textbf{PSNR} & \textbf{SSIM} \\
    \midrule
    Agriculture & \underline{24.6578} & \textbf{0.4472} & \textbf{24.8132} & \underline{0.4462} & 24.4998 & 0.4217 & 24.5131 & 0.4071 & 24.4603 & 0.3951 & 24.5596 & 0.4343 \\
Airplane & 27.4974 & 0.7942 & 27.6546 & 0.7989 & \textbf{27.8773} & \textbf{0.8047} & 27.5071 & 0.7965 & \underline{27.7454} & \underline{0.8013} & \underline{27.8502} & 0.8031 \\
Baseball Diamond & 32.2940 & 0.8262 & 32.4303 & 0.8275 & \underline{32.4722} & 0.8278 & 32.3994 & 0.8273 & 32.4668 & \underline{0.8287} & \textbf{32.5223} & \textbf{0.8297} \\
Beach & 35.1487 & 0.8817 & 35.6360 & 0.8818 & 35.5198 & \underline{0.8840} & 35.8321 & 0.8828 & \underline{35.8588} & \underline{0.8840} & \textbf{35.9922} & \textbf{0.8852} \\
Buildings & 23.5903 & 0.7311 & 23.6834 & 0.7379 & \textbf{23.8856} & \textbf{0.7473} & 23.6595 & 0.7361 & 23.8610 & \underline{0.7429} & \underline{23.8849} & 0.7447 \\
Chaparral & 25.3156 & 0.7439 & 25.3065 & 0.7448 & 25.3370 & 0.7456 & 25.2843 & 0.7431 & \underline{25.3508} & \underline{0.7473} & \textbf{25.4125} & \textbf{0.7487} \\
Dense Residential & 25.4557 & 0.7685 & 25.6661 & 0.7781 & \textbf{25.8694} & \textbf{0.7861} & 25.5968 & 0.7761 & 25.7106 & \underline{0.7800} & \underline{25.8308} & 0.7839 \\
Forest & \textbf{25.7158} & 0.6142 & 25.7046 & 0.6161 & 25.7154 & 0.6189 & 25.6836 & 0.6157 & 25.7118 & \underline{0.6201} & \underline{25.7200} & \textbf{0.6207} \\
Freeway & 27.0159 & 0.7241 & 27.2362 & \underline{0.7353} & \textbf{27.3893} & \textbf{0.7389} & 27.0942 & 0.7287 & \underline{27.2538} & 0.7347 & 27.2819 & 0.7347 \\
Golf Course & 32.6963 & 0.8288 & 32.8497 & 0.8298 & \textbf{32.9371} & 0.8318 & 32.8535 & 0.8303 & 32.8384 & \underline{0.8312} & \underline{32.9211} & \textbf{0.8324} \\
Harbor & 21.1856 & 0.8145 & 21.3915 & 0.8229 & \underline{21.4573} & \underline{0.8322} & 21.3021 & 0.8239 & 21.4672 & \textbf{0.8303} & \textbf{21.5047} & 0.8293 \\
Intersection & 25.3349 & 0.7334 & 25.4491 & 0.7408 & \underline{25.5327} & 0.7447 & 25.3997 & 0.7359 & \textbf{25.5408} & \underline{0.7446} & 25.5231 & \textbf{0.7451} \\
Medium Residential & 27.2653 & 0.7865 & 27.4423 & 0.7909 & \textbf{27.5991} & \textbf{0.7962} & 27.3904 & 0.7909 & 27.5026 & \underline{0.7933} & \underline{27.5707} & 0.7954 \\
Mobile Home Park & 17.4614 & 0.5832 & 17.4784 & 0.5916 & \underline{17.6597} & \textbf{0.6074} & 17.5639 & 0.5950 & 17.6232 & \underline{0.6007} & \textbf{17.6725} & 0.6011 \\
Overpass & 24.4138 & 0.6674 & 24.7141 & 0.6773 & \textbf{24.9631} & \textbf{0.6846} & 24.5785 & 0.6696 & 24.6946 & \underline{0.6765} & \underline{24.7885} & 0.6813 \\
Parking Lot & 21.1305 & 0.7261 & 21.2549 & 0.7373 & \underline{21.4778} & 0.7467 & 21.1929 & 0.7334 & 21.3833 & \underline{0.7420} & \textbf{21.4869} & \textbf{0.7473} \\
River & 29.0140 & 0.7675 & 29.0277 & 0.7670 & \underline{29.0942} & \underline{0.7710} & 28.9857 & 0.7678 & 29.0766 & 0.7714 & \textbf{29.1122} & \textbf{0.7718} \\
Runway & 29.6905 & 0.7382 & 29.8222 & 0.7490 & \textbf{30.1342} & \textbf{0.7563} & 29.7626 & 0.7395 & 29.8481 & \underline{0.7474} & \underline{30.0086} & 0.7533 \\
Sparse Residential & 27.0166 & 0.7537 & 27.1131 & 0.7563 & \textbf{27.2083} & \textbf{0.7606} & 27.0878 & 0.7547 & \underline{27.1766} & \underline{0.7604} & 27.1950 & 0.7598 \\
Storage Tanks & 28.8767 & 0.7977 & 29.0120 & 0.8015 & \textbf{29.1740} & \textbf{0.8061} & 29.0473 & 0.8014 & 29.0849 & \underline{0.8035} & \underline{29.1628} & 0.8056 \\
Tennis Court & 28.3825 & 0.7732 & 28.6161 & 0.7856 & \underline{28.7608} & 0.7928 & 28.4760 & 0.7759 & 28.6868 & \underline{0.7886} & \textbf{28.7917} & \textbf{0.7936} \\
\midrule
Average & 26.6266 & 0.7382 & 26.7763 & 0.7436 & \underline{26.8840} & \textbf{0.7479} & 26.7243 & 0.7396 & 26.8258 & \underline{0.7440} & \textbf{26.8949} & 0.7477 \\
    \bottomrule
  \end{tabular}
  }
  \label{tab:ucmerced_result}
\end{table*}

\begin{table*}[pos=t]
  \centering
  \caption{Quantitative Results of the experiment on the AID. The best result is highlighted in \textbf{bold}, while the second-best result is \underline{underlined}. The unit for PSNR is decibels (dB).}
  \resizebox{\textwidth}{!}{%
  \begin{tabular}{ccccccccccccc}
    \toprule
    \multirow{3}{*}{\textbf{Class}} 
    & \multicolumn{2}{c}{\textbf{FAT}}
    & \multicolumn{2}{c}{\textbf{FMSR}}
    & \multicolumn{2}{c}{\textbf{GCRDN}}
    & \multicolumn{2}{c}{\textbf{TSFNet}}
    & \multicolumn{2}{c}{\textbf{TTST}} 
    & \multicolumn{2}{c}{\textbf{Text-RSIR}} \\
    \cmidrule(lr){2-3} \cmidrule(lr){4-5} \cmidrule(lr){6-7} \cmidrule(lr){8-9} \cmidrule(lr){10-11} \cmidrule(lr){12-13}
    & \textbf{PSNR} & \textbf{SSIM} 
    & \textbf{PSNR} & \textbf{SSIM} 
    & \textbf{PSNR} & \textbf{SSIM} 
    & \textbf{PSNR} & \textbf{SSIM} 
    & \textbf{PSNR} & \textbf{SSIM} 
    & \textbf{PSNR} & \textbf{SSIM} \\
    \midrule
    Airport & 27.7525 & 0.7324 & \underline{27.9582} & \textbf{0.7419} & 27.6506 & 0.7277 & 27.7524 & 0.7337 & 27.9100 & 0.7384 & \textbf{27.9673} & \underline{0.7401} \\
    Bare Land & 34.5639 & 0.8269 & 34.7455 & \underline{0.8323} & 34.6666 & 0.8289 & 34.6324 & 0.8304 & \underline{34.7469} & 0.8322 & \textbf{34.8499} & \textbf{0.8328} \\
    Baseball Field & 27.7234 & 0.7363 & \textbf{27.9125} & \textbf{0.7459} & 27.6471 & 0.7350 & 27.7408 & 0.7401 & 27.7956 & 0.7427 & \underline{27.9062} & \underline{0.7454} \\
    Beach & 31.5206 & 0.7959 & 31.7368 & 0.8017 & 31.6311 & 0.7967 & 31.6762 & 0.7993 & \underline{31.7811} & \underline{0.8025} & \textbf{31.8627} & \textbf{0.8028} \\
    Bridge & 29.5164 & 0.8066 & \underline{29.7611} & \textbf{0.8140} & 29.3744 & 0.8033 & 29.5317 & 0.8083 & 29.7168 & 0.8122 & \textbf{29.7696} & \underline{0.8134} \\
    Center & 24.1964 & 0.6563 & \underline{24.4127} & \textbf{0.6683} & 24.0441 & 0.6476 & 24.2564 & 0.6598 & 24.3907 & \underline{0.6659} & \textbf{24.4493} & \textbf{0.6683} \\
    Church & 23.5623 & 0.6633 & \underline{23.7468} & \textbf{0.6747} & 23.4374 & 0.6554 & 23.5907 & 0.6661 & 23.7075 & 0.6711 & \textbf{23.7721} & \underline{0.6743} \\
    Commercial & 24.3682 & 0.6655 & \underline{24.5198} & \textbf{0.6741} & 24.2827 & 0.6595 & 24.4117 & 0.6687 & 24.4918 & 0.6717 & \textbf{24.5252} & \underline{0.6737} \\
    Dense Residential & 25.4483 & 0.6691 & \textbf{25.5893} & \textbf{0.6774} & 25.3104 & 0.6597 & 25.4380 & 0.6699 & 25.5417 & 0.6724 & \underline{25.5836} & \underline{0.6755} \\
    Desert & 35.4974 & 0.8603 & \underline{35.6769} & 0.8646 & 35.5656 & 0.8623 & 35.4717 & 0.8622 & 35.5750 & \underline{0.8649} & \textbf{35.8370} & \textbf{0.8652} \\
    Farmland & 32.3457 & 0.8001 & \textbf{32.5888} & \textbf{0.8090} & 32.2252 & 0.7972 & 32.2969 & 0.8011 & 32.4639 & 0.8054 & \underline{32.5196} & \underline{0.8064} \\
    Forest & 25.6715 & 0.5934 & \textbf{25.8384} & \underline{0.5994} & 25.6358 & 0.5859 & 25.7607 & 0.5953 & 25.8171 & 0.5971 & \textbf{25.8607} & \textbf{0.5998} \\
    Industrial & 25.3502 & 0.6885 & \underline{25.5426} & \textbf{0.6982} & 25.1632 & 0.6782 & 25.3566 & 0.6899 & 25.4803 & 0.6942 & \textbf{25.5438} & \underline{0.6970} \\
    Meadow & 30.5650 & 0.6421 & \underline{30.7640} & \textbf{0.6548} & 30.6964 & 0.6493 & 30.7205 & 0.6529 & 30.7427 & 0.6536 & \textbf{30.7743} & \underline{0.6544} \\
    Medium Residential & 24.1298 & 0.5843 & \textbf{24.3298} & \underline{0.5951} & 24.0050 & 0.5725 & 24.2174 & 0.5894 & 24.2783 & 0.5911 & \underline{24.3296} & \textbf{0.5952} \\
    Mountain & 26.9741 & 0.6309 & 27.0518 & 0.6337 & 26.9325 & 0.6250 & 26.9786 & 0.6326 & \underline{27.0588} & \underline{0.6346} & \textbf{27.0870} & \textbf{0.6357} \\
    Park & 25.2941 & 0.6589 & 25.4161 & \underline{0.6655} & 25.2218 & 0.6533 & 25.3075 & 0.6610 & 25.4103 & 0.6638 & \textbf{25.4422} & \textbf{0.6663} \\
    Parking & 24.5978 & 0.7433 & \textbf{24.9916} & \textbf{0.7607} & 24.1571 & 0.7263 & 24.6874 & 0.7474 & 24.3829 & 0.7408 & \underline{24.8294} & \underline{0.7539} \\
    Playground & 30.0128 & 0.7827 & \underline{30.2509} & \textbf{0.7918} & 29.7851 & 0.7770 & 30.0155 & 0.7848 & 30.1723 & 0.7888 & \textbf{30.2731} & \underline{0.7910} \\
    Pond & 27.8899 & 0.7152 & \underline{27.9814} & \underline{0.7191} & 27.8211 & 0.7123 & 27.8758 & 0.7161 & 27.9796 & 0.7186 & \textbf{28.0043} & \textbf{0.7200} \\
    Port & 26.2843 & 0.7471 & \underline{26.4492} & \textbf{0.7545} & 26.1971 & 0.7440 & 26.2961 & 0.7489 & 26.4259 & 0.7528 & \textbf{26.4616} & \underline{0.7541} \\
    Railway Station & 25.5554 & 0.6620 & \underline{25.7163} & \textbf{0.6698} & 25.3999 & 0.6506 & 25.5058 & 0.6609 & 25.6945 & 0.6667 & \textbf{25.7389} & \underline{0.6695} \\
    Resort & 22.7878 & 0.6255 & \underline{22.9452} & \underline{0.6342} & 22.7181 & 0.6206 & 22.7916 & 0.6278 & 22.9095 & 0.6325 & \textbf{22.9499} & \textbf{0.6345} \\
    River & 28.5470 & 0.6859 & \underline{28.6762} & \underline{0.6926} & 28.5350 & 0.6848 & 28.5804 & 0.6897 & 28.6669 & 0.6919 & \textbf{28.6964} & \textbf{0.6928} \\
    School & 24.0299 & 0.6776 & \underline{24.2210} & \underline{0.6871} & 23.9095 & 0.6716 & 24.0555 & 0.6810 & 24.1896 & 0.6849 & \textbf{24.2368} & \textbf{0.6876} \\
    Sparse Residential & 24.6960 & 0.5454 & \underline{24.8526} & \textbf{0.5526} & 24.6912 & 0.5397 & 24.7662 & 0.5477 & 24.8212 & 0.5498 & \textbf{24.8562} & \underline{0.5524} \\
    Square & 28.3506 & 0.7540 & \underline{28.5661} & \textbf{0.7636} & 28.1843 & 0.7483 & 28.3606 & 0.7565 & 28.5292 & 0.7613 & \textbf{28.5974} & \underline{0.7634} \\
    Stadium & 26.0934 & 0.7377 & \underline{26.3204} & \underline{0.7486} & 25.9347 & 0.7310 & 26.1205 & 0.7399 & 26.2661 & 0.7471 & \textbf{26.3626} & \textbf{0.7502} \\
    Storage Tanks & 24.8955 & 0.6788 & \textbf{25.0844} & \textbf{0.6894} & 24.7784 & 0.6727 & 24.9263 & 0.6815 & 25.0396 & 0.6863 & \underline{25.0742} & \underline{0.6881} \\
    Viaduct & 25.6030 & 0.6395 & \underline{25.7718} & \textbf{0.6497} & 25.4949 & 0.6314 & 25.6565 & 0.6427 & 25.7504 & \underline{0.6467} & \textbf{25.7897} & \textbf{0.6497} \\
    \midrule
    Average & 27.2062 & 0.7034 & \underline{27.3927} & \textbf{0.7120} & 27.1123 & 0.6981 & 27.2373 & 0.7060 & 27.3347 & 0.7092 & \textbf{27.4089} & \underline{0.7116} \\
    \bottomrule
  \end{tabular}
  }
  \label{tab:aid_result}
\end{table*}
\begin{table}[pos=t]
  \centering
  \caption{Quantitative Results of the experiment on the ILSVRC2012 dataset. 
  The best result is highlighted in \textbf{bold}, while the second-best result is \underline{underlined}.}
  \resizebox{\columnwidth}{!}{%
  \begin{tabular}{ccccccc}
    \toprule
    \textbf{Model} 
    & FAT & FMSR & GCRDN & TSFNet & TTST & Text-RSIR \\
    \midrule
    \textbf{PSNR} (dB) & 23.7829 & 24.0107 & 23.9232 & 23.8665 & \underline{24.0294} & \textbf{24.0528} \\
    \textbf{SSIM} & 0.6485 & 0.6585 & 0.6527 & 0.6489 & \underline{0.6591} & \textbf{0.6599} \\
    \bottomrule
  \end{tabular}
  }
  \label{tab:imagenet_result}
\end{table}

\begin{figure*}[pos=t]
\centering
\begin{subfigure}{0.47\textwidth}
  \centering
  \includegraphics[width=\linewidth]{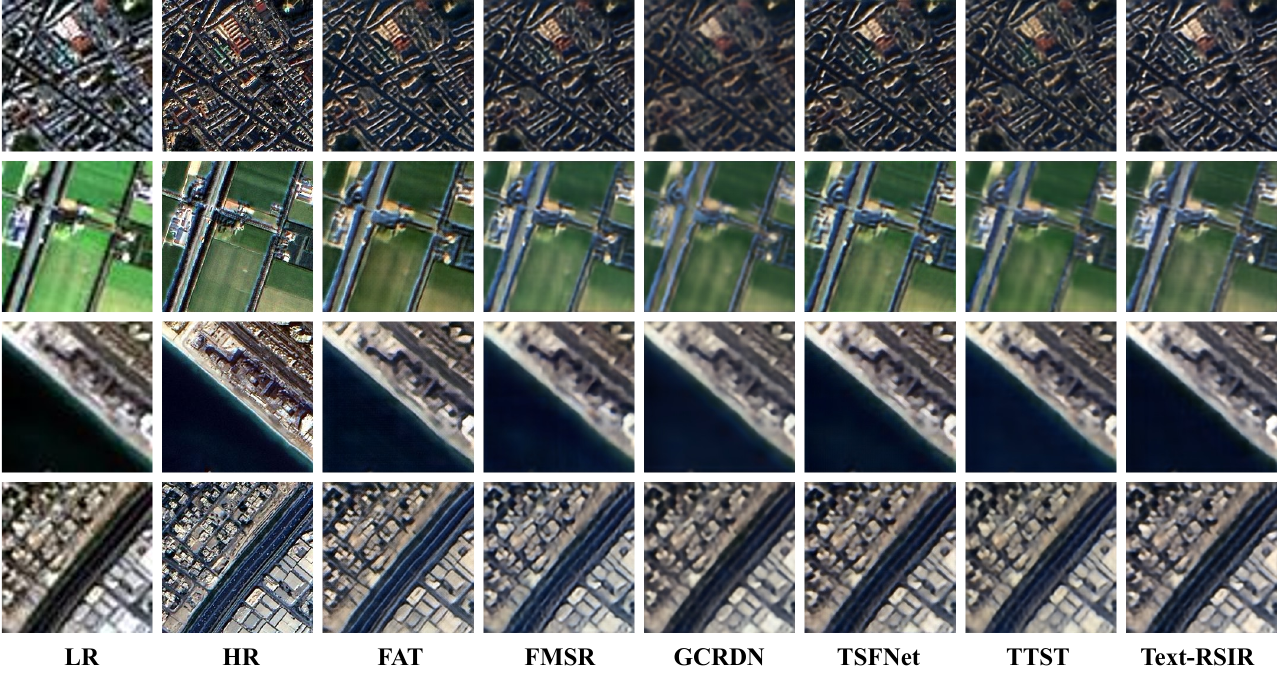}
  \caption{Qualitative results on Alsat-2B.}
  \label{fig:compare_alsat}
\end{subfigure}
\hfill
\begin{subfigure}{0.47\textwidth}
  \centering
  \includegraphics[width=\linewidth]{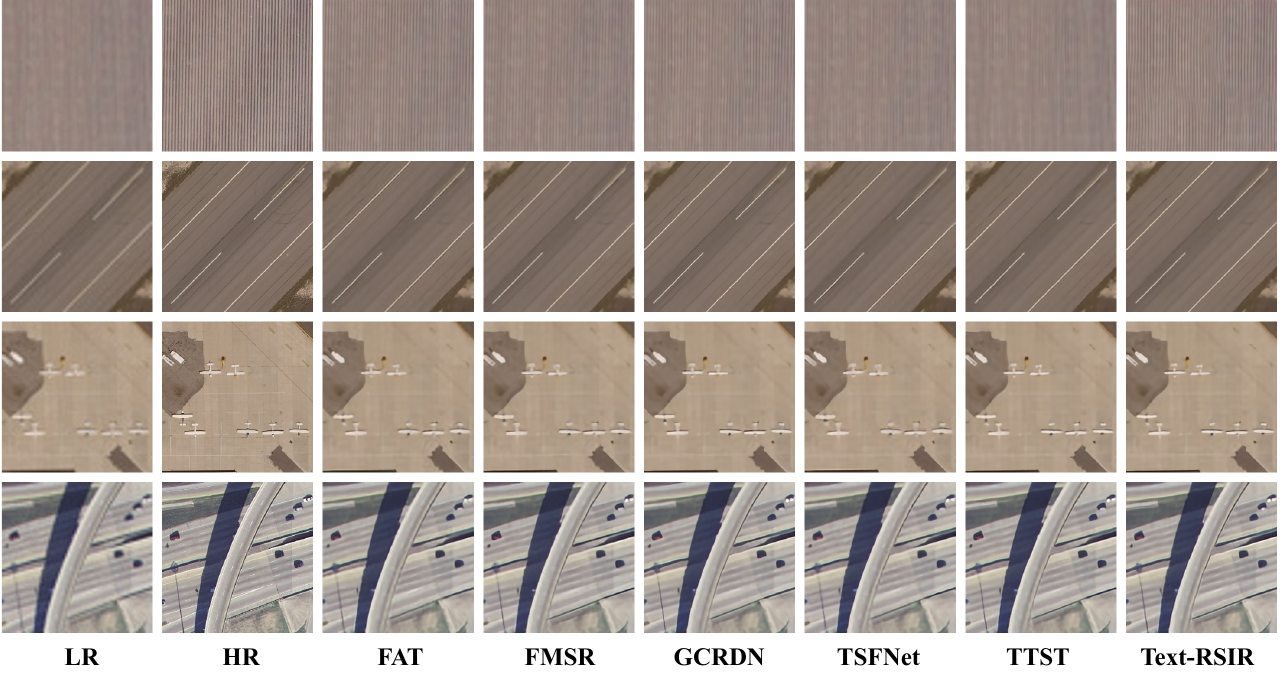}
  \caption{Qualitative results on UC Merced Land Use.}
  \label{fig:compare_ucmerced}
\end{subfigure}
\vspace{0.25em}
\begin{subfigure}{0.47\textwidth}
  \centering
  \includegraphics[width=\linewidth]{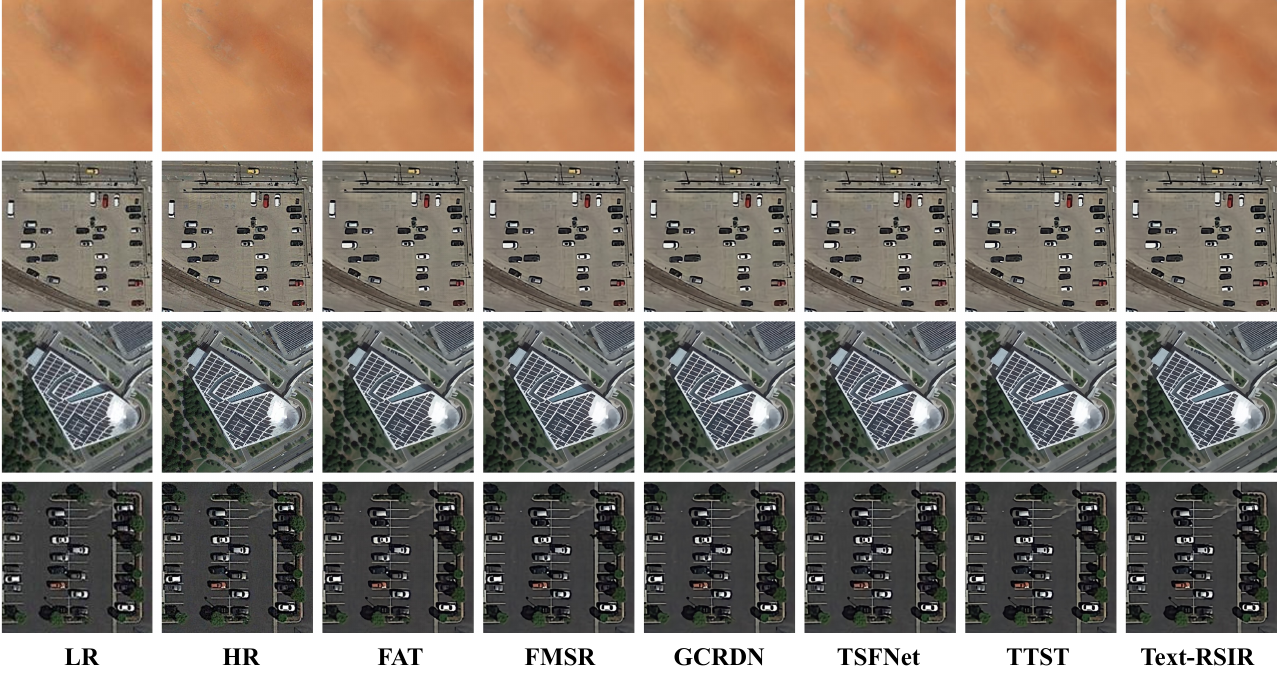}
  \caption{Qualitative results on AID.}
  \label{fig:compare_aid}
\end{subfigure}
\hfill
\begin{subfigure}{0.47\textwidth}
  \centering
  \includegraphics[width=\linewidth]{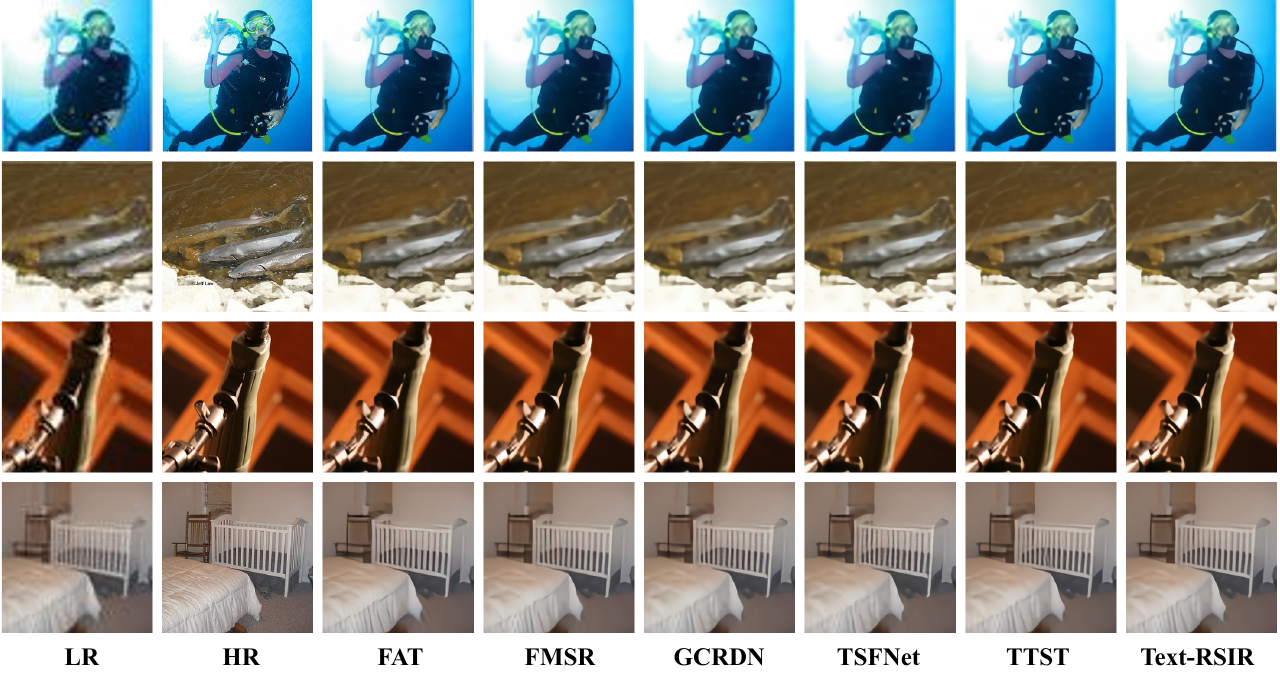}
  \caption{Qualitative results on ILSVRC2012.}
  \label{fig:compare_imagenet}
\end{subfigure}
\caption{Visualization of results from Text-RSIR and other methods across all four datasets. Subfigures are labeled (a)–(d).}
\label{fig:compare}
\end{figure*}

To further evaluate the visual quality of the SRIs, we compare the results of Text-RSIR with those of other models across all four datasets, as shown in Figure \ref{fig:compare}. 
In addition to achieving higher PSNR and SSIM scores, Text-RSIR also demonstrates superior visual fidelity. 
When a domain shift exists between LR and HR images, as illustrated in the first and second rows in Figure \ref{fig:compare}\subref{fig:compare_alsat}, Text-RSIR successfully restores the correct color domain of the HR images. 
In cases involving repetitive line patterns, as seen in the first, second and fourth rows in Figure \ref{fig:compare}\subref{fig:compare_ucmerced}, Text-RSIR reconstructs the lines directly rather than simply emphasizing them. Moreover, Text-RSIR performs particularly well on parking lot images, achieving state-of-the-art results compared with other methods, as shown in the second and fourth rows in Figure \ref{fig:compare}\subref{fig:compare_aid}. 
Lastly, Text-RSIR demonstrates superior performance in reconstructing common object images compared to other methods, as shown in Figure \ref{fig:compare}\subref{fig:compare_imagenet}. 
While it excels at recovering repeatable patterns, such as the railing of the baby bed in the last row, it also outperforms others in restoring contextual details, as seen in the reconstruction of the bell in the third row.

\begin{figure*}[pos=t]
\centering
\begin{subfigure}{0.47\textwidth}
  \centering
  \includegraphics[width=\linewidth]{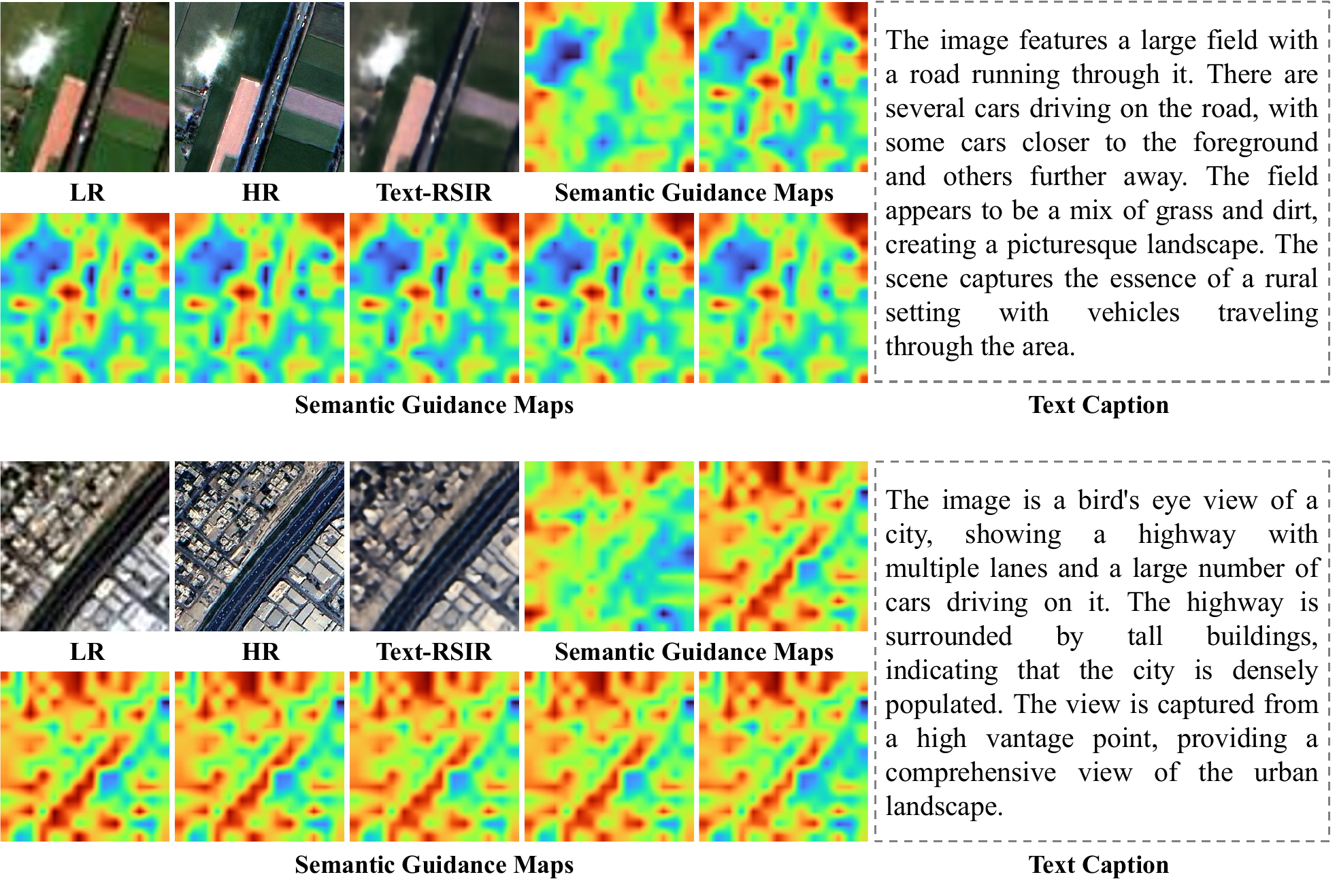}
  \caption{Qualitative results on Alsat-2B.}
  \label{fig:vis_alsat}
\end{subfigure}
\hfill
\begin{subfigure}{0.47\textwidth}
  \centering
  \includegraphics[width=\linewidth]{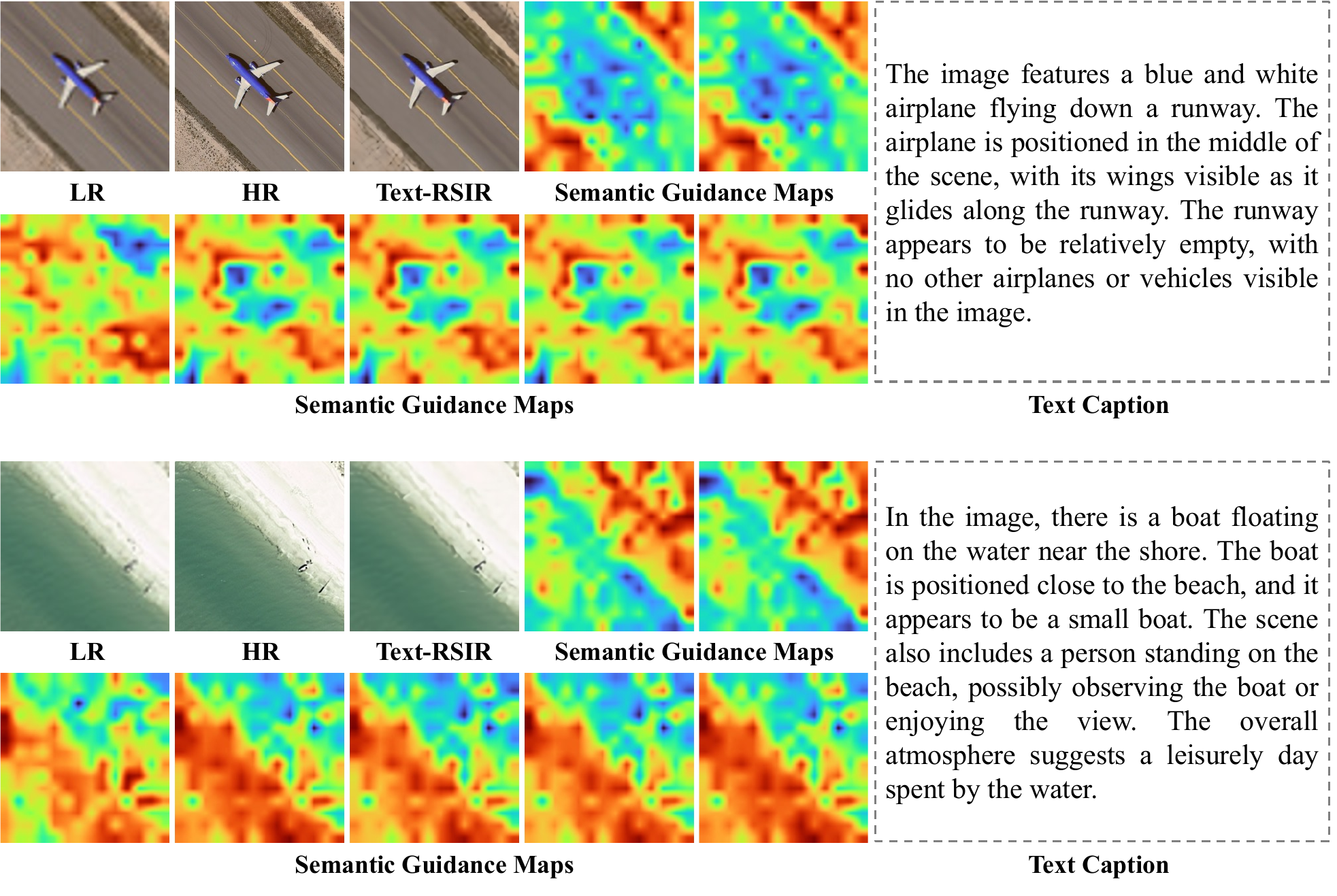}
  \caption{Qualitative results on UC Merced Land Use.}
  \label{fig:vis_ucmerced}
\end{subfigure}
\vspace{0.25em}
\begin{subfigure}{0.47\textwidth}
  \centering
  \includegraphics[width=\linewidth]{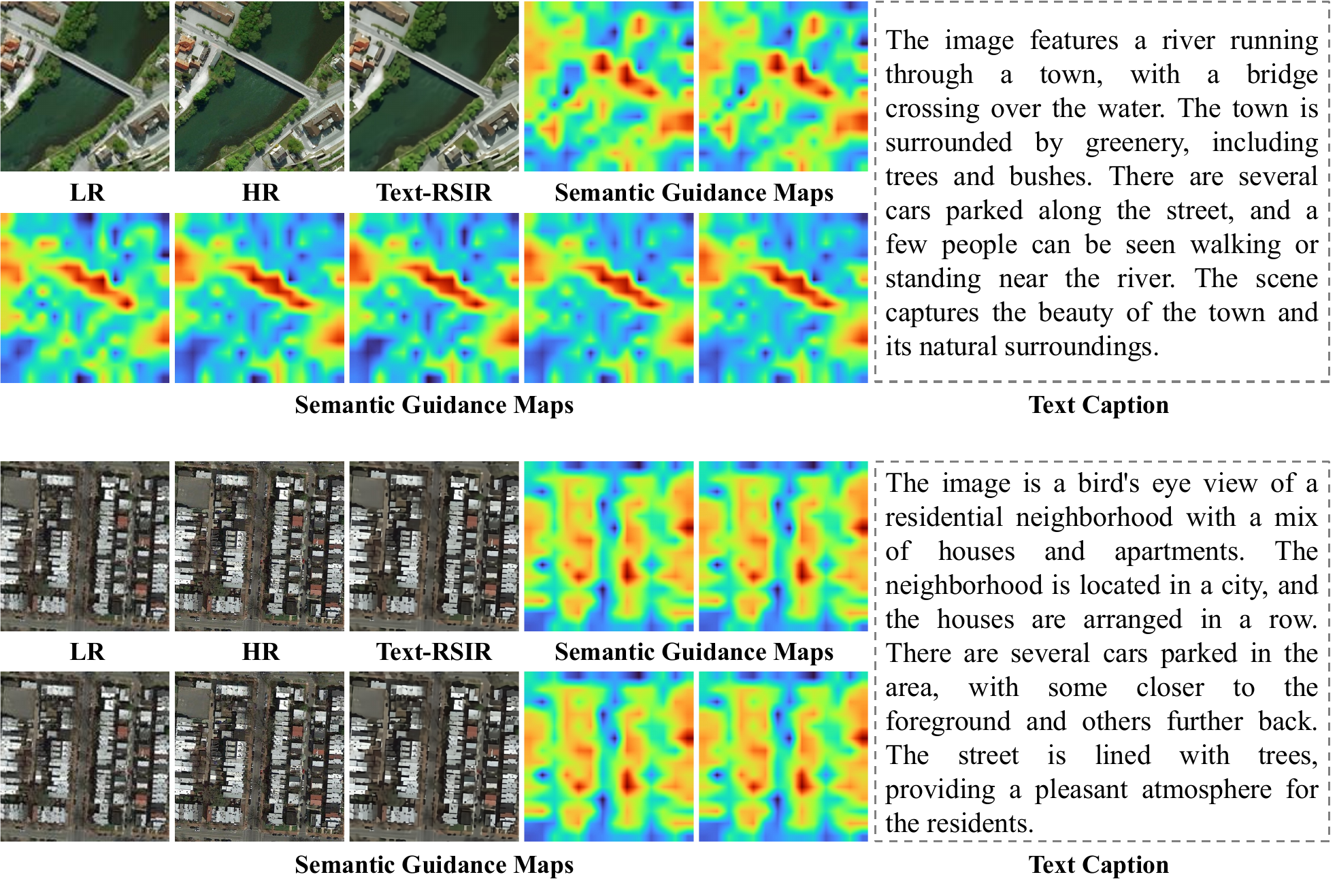}
  \caption{Qualitative results on AID.}
  \label{fig:vis_aid}
\end{subfigure}
\hfill
\begin{subfigure}{0.47\textwidth}
  \centering
  \includegraphics[width=\linewidth]{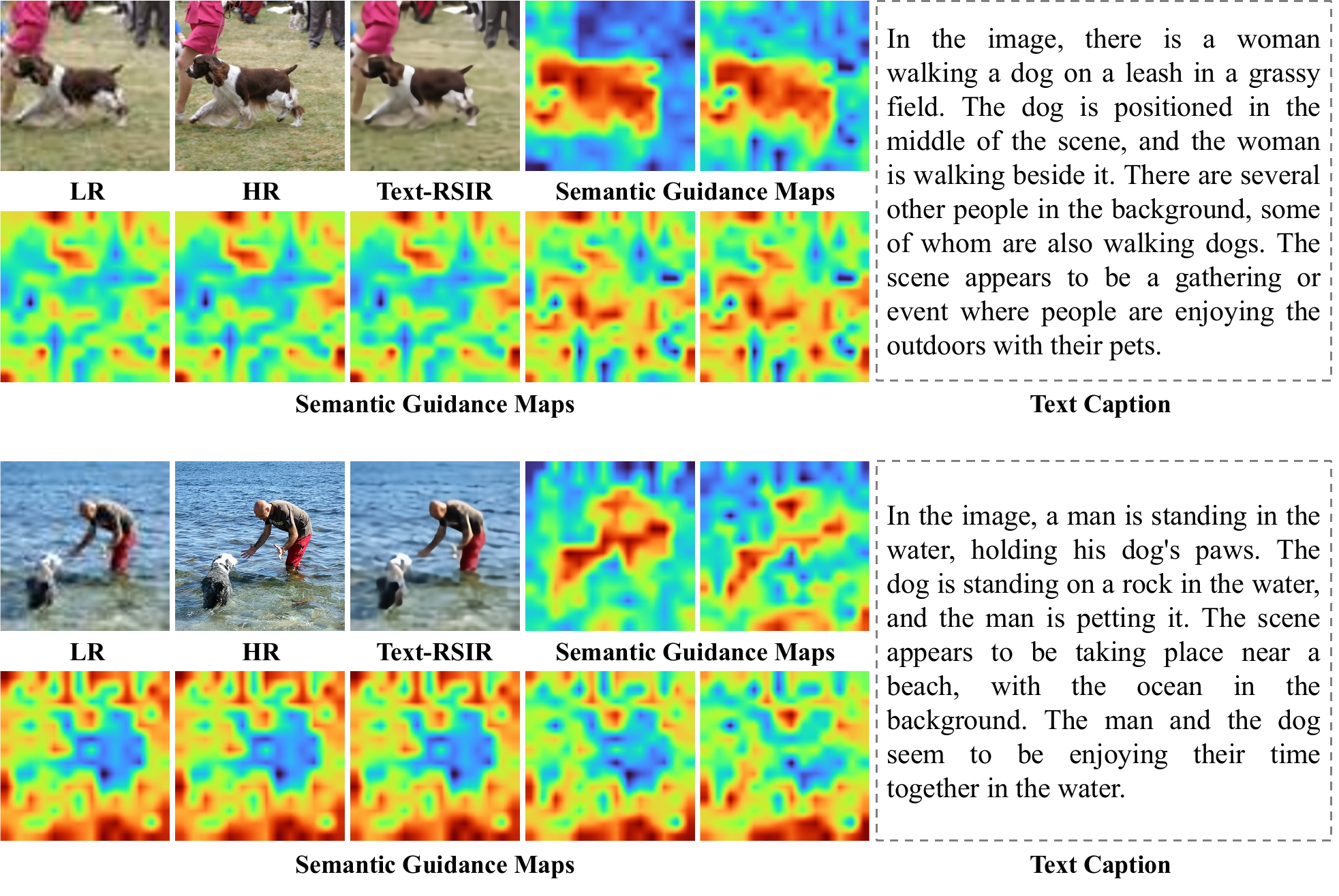}
  \caption{Qualitative results on ILSVRC2012.}
  \label{fig:vis_imagenet}
\end{subfigure}
\caption{Visualization of Text-RSIR results across all four datasets, including both the TCs and heatmap visualizations of the SGMs. The TCs often provide additional details that may be overlooked by the SGMs, while the SGMs from different iterations in TGISR often attend to different image regions. Subfigures are labeled (a)–(d).}
\label{fig:vis}
\end{figure*}

We also present a visualization of Text-RSIR results across all four datasets in Figure \ref{fig:vis}, which includes both TCs and heatmap visualizations of the SGMs. 
The TC often provide additional details that may be overlooked by the SGMs, thereby contributing missing information during the RSSR process. 
For instance, in the first sample in Figure~\ref{fig:vis}\subref{fig:vis_alsat}, the TC refers to roads and cars, and in the first sample in Figure~\ref{fig:vis}\subref{fig:vis_aid}, the TC refers to trees and bushes, which are not captured by the SGMs.
The figure further illustrates that the model exhibits varying focus at different stages of the TGISR process. 
For example, in the second sample of Figure~\ref{fig:vis}\subref{fig:vis_ucmerced}, the model emphasizes the sea during the third, fourth, and fifth iterations, while focusing on the beach and high-frequency contextual details in other iterations. 
In the second sample of Figure~\ref{fig:vis}\subref{fig:vis_imagenet}, the model emphasizes the man during the first two iterations, while focusing on the background items in the other iterations. 
In summary, both TCs generated by the VLM and the corresponding SGMs contribute meaningfully to HR image reconstruction.

Since TC provides a detailed description of the image, incorporating it into the SR process may enhance the domain adaptability of Text-RSIR. 
To further evaluate that ability, we conduct a domain adaptation experiment between the Alsat-2B dataset and the UC Merced Land Use Dataset. 
Additionally, we perform another experiment in which models are trained on the ILSVRC2012 dataset and tested on both the Alsat-2B and the UC Merced Land Use datasets. 
The quantitative results are summarized in Table \ref{tab:da}. 
To enable a fairer comparison with the other datasets, compression is not applied to the UC Merced Land Use Dataset.
These results indicate that Text-RSIR exhibits strong domain adaptability. Specifically, it achieves the highest SSIM index (0.5364) when trained on the Alsat-2B dataset and tested on the UC Merced Land Use Dataset. 
It also attains the second-best PSNR value (13.5755 dB) when trained on the UC Merced dataset and tested on Alsat-2B. 
Moreover, when trained on common object images, Text-RSIR achieves the best performance in both PSNR and SSIM when evaluated on the UC Merced dataset, which contains a broader variety of scene types than the Alsat-2B dataset. 
Overall, these findings confirm that Text-RSIR is highly adaptable across domains, and its performance can be further enhanced by training on datasets with greater scene diversity.

\begin{table*}[pos=t]
  \centering
  \caption{Quantitative results of the domain adaptation experiments, with all datasets abbreviated using their initial capital letter. The training dataset is listed to the left of the \textrightarrow, and the testing dataset to the right. The best result is highlighted in \textbf{bold}, while the second-best result is \underline{underlined}. The unit for PSNR is decibels (dB).}
  \resizebox{\textwidth}{!}{%
  \begin{tabular}{ccccccccc}
    \toprule
    \multirow{3}{*}{\textbf{Model}} 
    & \multicolumn{2}{c}{\textbf{A. \textrightarrow U.}} 
    & \multicolumn{2}{c}{\textbf{U. \textrightarrow A.}} 
    & \multicolumn{2}{c}{\textbf{I. \textrightarrow A.}} 
    & \multicolumn{2}{c}{\textbf{I. \textrightarrow U.}} \\
    \cmidrule(lr){2-3} \cmidrule(lr){4-5} \cmidrule(lr){6-7} \cmidrule(lr){8-9}
    & \textbf{PSNR} & \textbf{SSIM} & \textbf{PSNR} & \textbf{SSIM} & \textbf{PSNR} & \textbf{SSIM} & \textbf{PSNR} & \textbf{SSIM} \\
    \midrule
    FAT & 19.2788 & 0.4559 & 13.5311 & 0.2497 & 13.6382 & 0.2705 & 25.6201 & 0.6860 \\
    FMSR & \underline{19.9059} & 0.5316 & 13.5659 & 0.2518 & 13.6966 & 0.2756 & 25.8744 & \underline{0.6938} \\
    GCRDN & \textbf{20.9480} & \underline{0.5356} & 13.5551 & \underline{0.2556} & \underline{13.7073} & \underline{0.2770} & \underline{25.8940} & 0.6934\\
    TSFNet & 19.6710 & 0.4932 & 13.5017 & 0.2363 & \textbf{13.7622} & \textbf{0.2782} & 25.7703 & 0.6873\\
    TTST & 19.2694 & 0.5094 & \textbf{13.6064} & \textbf{0.2559} & 13.6952 & 0.2754 & 25.7856 & 0.6906\\
    Text-RSIR & 19.7900 & \textbf{0.5364} & \underline{13.5755} & 0.2552 & 13.6798 & 0.2753 & \textbf{25.8997} & \textbf{0.6947} \\
    \bottomrule
  \end{tabular}
  }
  \label{tab:da}
\end{table*}

\begin{figure}[pos=t]
\centering
\includegraphics[width=\columnwidth]{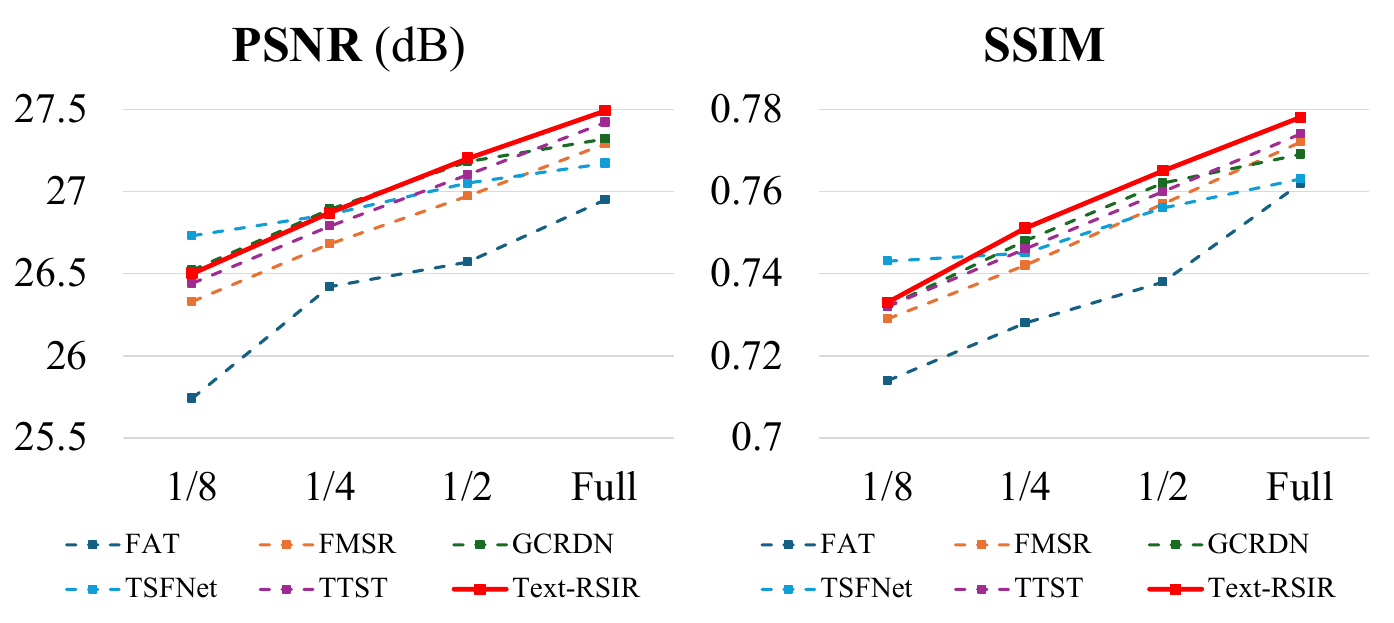}
\caption{Visualization of the data scalability experiment results on the UC Merced Land Use Dataset.}
\label{fig:ds}
\end{figure}

With the additional TC input compared to other methods, Text-RSIR has access to more information and may demonstrate superior performance in scenarios with limited training data. 
To further assess the scalability of Text-RSIR in that scenario, we conduct a data scalability experiment on the UC Merced Land Use Dataset.
Specifically, we construct three reduced training sets containing only 1/2, 1/4, and 1/8 of the original training images, respectively, with no compression applied to ensure a clear and consistent comparison.
The quantitative results are shown in Table \ref{tab:ds}, and the visualization of the results is shown in Figure \ref{fig:ds}. 
The results indicate that Text-RSIR exhibits the fastest performance growth as the training set size increases, while still maintaining competitive results with very limited data. 
Specifically, Text-RSIR achieves state-of-the-art performance when trained on half and the full partition of the training data. It also attains the second-best SSIM when trained on 1/8 of the data, and ranks second in PSNR and first in SSIM when trained on 1/4 of the data. 
This suggests that the incorporation of text guidance not only improves reconstruction quality with full training data but also provides strong robustness in low-data regimes, highlighting the model’s generalization capability and practical utility in scenarios where annotated data is scarce.

\begin{table*}[pos=t]
  \centering
  \caption{Quantitative results of the data scalability experiments on the UC Merced Land Use Dataset. “1/8”, “1/4”, “1/2”, and “Full” indicate the proportion of the training set used. The best result is highlighted in \textbf{bold}, while the second-best result is \underline{underlined}. The unit for PSNR is decibels (dB).}
  \resizebox{\textwidth}{!}{%
  \begin{tabular}{ccccccccc}
    \toprule
    \multirow{3}{*}{\textbf{Model}} 
    & \multicolumn{2}{c}{\textbf{1/8}} 
    & \multicolumn{2}{c}{\textbf{1/4}} 
    & \multicolumn{2}{c}{\textbf{1/2}} 
    & \multicolumn{2}{c}{\textbf{Full}} \\
    \cmidrule(lr){2-3} \cmidrule(lr){4-5} \cmidrule(lr){6-7} \cmidrule(lr){8-9}
    & \textbf{PSNR} & \textbf{SSIM} & \textbf{PSNR} & \textbf{SSIM} & \textbf{PSNR} & \textbf{SSIM} & \textbf{PSNR} & \textbf{SSIM} \\
    \midrule
    FAT & 25.7443 & 0.7137 & 26.4231 & 0.7280 & 26.5719 & 0.7376 & 26.9460 & 0.7617 \\
    FMSR & 26.3315 & 0.7290 & 26.6758 & 0.7418 & 26.9741 & 0.7571 & 27.2922 & 0.7718 \\
    GCRDN & \underline{26.5177} & 0.7319 & \textbf{26.8927} & \underline{0.7484} & \underline{27.1808} & \underline{0.7620} & 27.3259 & 0.7688\\
    TSFNet & \textbf{26.7270} & \textbf{0.7428} & 26.8622 & 0.7475 & 27.0533 & 0.7563 & 27.1664 & 0.7631 \\
    TTST & 26.4420 & 0.7324 & 26.7914 & 0.7461 & 27.1010 & 0.7603 & \underline{27.4235} & \underline{0.7738} \\
    Text-RSIR & 26.5027 & \underline{0.7331} & \underline{26.8675} & \textbf{0.7512} & \textbf{27.1958} & \textbf{0.7648} & \textbf{27.4904} & \textbf{0.7775}\\
    \bottomrule
  \end{tabular}
  }
  \label{tab:ds}
\end{table*}

\subsection{Ablation Study}

To further evaluate the effectiveness of each components of TGFA process, this study conducts an ablation study on TGFA on the UC Merced Land Use Dataset. 
We first construct a model with a single TGFA process applied before the first SR Unit, using only SGMs and removing the TE. 
Next, we build a second model that uses only TE, excluding the SGMs. 
We then create a third model that incorporates both SGMs and TE. 
Subsequently, we design a hierarchical architecture by adding a TGFA process after each SR Unit. 
Finally, we introduce a learnable aligning factor for each TGFA process to fine-tune the alignment, resulting in the complete Text-RSIR model.

The quantitative results of the ablation study are presented in Table \ref{tab:ablation}. 
To ensure a clear and consistent comparison, no compression is applied.
As shown, incorporating SGMs as pixel-wise guidance or a TE as channel-wise guidance both enhance performance, with TE yielding a slightly greater improvement (0.0042 dB in PSNR and 0.0018 in SSIM) compared to SGMs. 
Combining both types of embeddings further boosts reconstruction quality, demonstrating that pixel-wise and channel-wise guidance provide complementary information for the feature aligning process. 
Introducing the hierarchical architecture brings an additional performance gain (0.0191 dB in PSNR and 0.0023 in SSIM), indicating that multi-level alignment enhances the integration of visual and textual cues. 
Finally, adding a learnable aligning factor provides the Text-RSIR model with greater flexibility in leveraging text guidance, leading to a further improvement in performance.

\begin{table}[pos=t]
  \centering
  \caption{Quantitative results of the ablation study on TGFA for the UC Merced Land Use Dataset. \textit{SGMs} refers to Semantic Guidance Maps, \textit{TE} to Textual Embeddings, \textit{Hierarchy} to the hierarchical architecture of the TGFA process, and \textit{LAF} to the learnable aligning factor. The best result is shown in \textbf{bold}, while the second-best result is \underline{underlined}.}
  \begin{tabular}{cccccc}
    \toprule
    \multicolumn{4}{c}{\textbf{Model Settings}}
    & \multicolumn{2}{c}{\textbf{Metrics}} \\
    \cmidrule(lr){1-4} \cmidrule(lr){5-6} 
    SGMs & TE & Hierarchy & LAF & \textbf{PSNR} (dB) & \textbf{SSIM} \\
    \midrule
    $\checkmark$ & & & & 27.4464 & 0.7726\\
    & $\checkmark$ & & & 27.4506 & 0.7744\\
    $\checkmark$ & $\checkmark$ & & & 27.4580 & 0.7750\\
    $\checkmark$ & $\checkmark$ & $\checkmark$ &  & \underline{27.4771} & \underline{0.7773}\\
    $\checkmark$ & $\checkmark$ & $\checkmark$ & $\checkmark$ & \textbf{27.4904} & \textbf{0.7775}\\
    \bottomrule
  \end{tabular}
  \label{tab:ablation}
\end{table}

\begin{figure}[pos=t]
\centering
\includegraphics[width=\columnwidth]{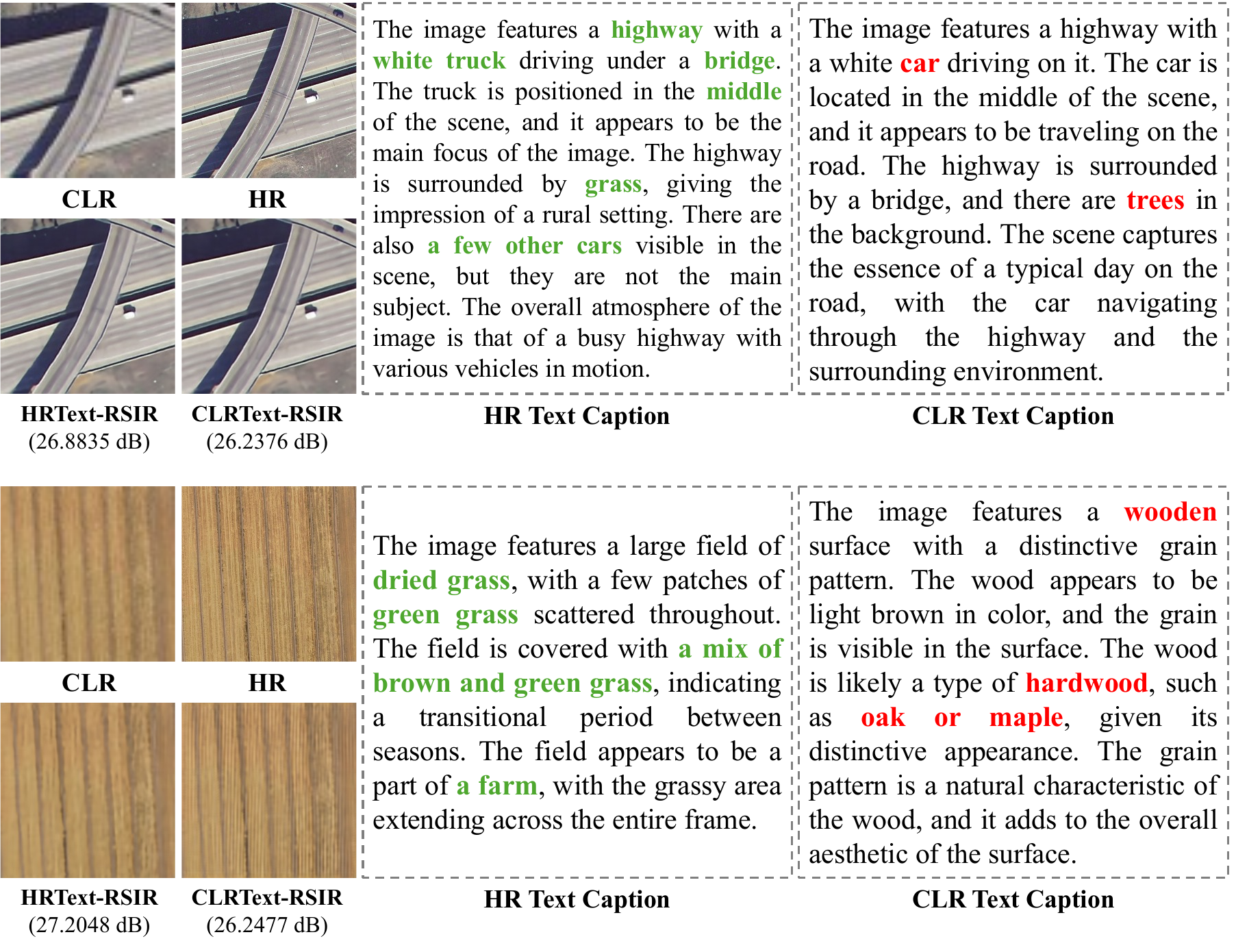}
\caption{Qualitative comparison between the HR-generated and CLR-generated TCs. Super-resolved images generated using HR TCs are labeled as "HRText-RSIR," while those generated using CLR TCs are labeled as "CLRText-RSIR." The PSNR values between the super-resolved images and the HR ground truth are displayed below each label. In the TCs, words and phrases that accurately describe the HR images are highlighted in \textbf{bold} green, whereas those that are incorrect or irrelevant are highlighted in \textbf{bold} red.}
\label{fig:lrt}
\end{figure}

To further evaluate the effectiveness of transferring TCs from HR images, this study conducts an ablation experiment in which CLR-generated captions replace the original HR-generated captions on the UC Merced Land Use Dataset.
In this setting, Text-RSIR achieves a PSNR of 26.8588 dB and an SSIM of 0.7475. 
A qualitative comparison between HR- and CLR-generated TCs is shown in Figure \ref{fig:lrt}. 
The results indicate that in some cases, text degradation can lead to a PSNR drop of approximately 1 dB, highlighting the importance of transferring HR-generated TCs from the satellite to the ground. 
Incorrect contextual descriptions in the captions (\textit{e.g.}, “trees” in the first sample and “hardwood” in the second) can mislead the model and degrade SR performance. 
Additionally, generating captions from CLR images provides the VLM with less visual information, resulting in less informative and more misleading TCs.
However, since the performance degradation is minimal, Text-RSIR also demonstrates robustness to the quality of the TCs, indicating that more lightweight text generators remain viable options.

\section{Conclusion}

This study has proposed TGRSIT, a text-guided remote sensing image transmission system that downlinks lightweight LR–text pairs, together with Text-RSIR as the ground-side text-conditioned reconstruction network.
By shifting from conventional pixel-intensive HR downlink to LR-text transmission, TGRSIT substantially reduces communication overhead while preserving task-relevant contextual information through onboard VLM-generated textual captions.
On the receiver side, Text-RSIR exploits the transmitted captions and CLIP-derived embeddings to provide complementary semantic priors beyond the LR image content.
Through iterative text-guided feature alignment and refinement, the network effectively enhances fine-grained spatial detail recovery while maintaining global semantic consistency.
Experimental results have validated both the effectiveness of the proposed TGRSIT and the reconstruction capability of Text-RSIR.
TGRSIT achieves a reconstruct PSNR of 26.87 and an SSIM of 0.7475, while keeping the downlink size to around 2\% as large the full-pixel dataset.
Text-RSIR has consistently outperformed state-of-the-art methods in both PSNR and SSIM across Alsat-2B, UC Merced Land Use, Aerial Image, and ILSVRC2012 datasets, while maintaining competitive computational efficiency. 
The ablation study has further confirmed that the hierarchical integration of CLIP visual and textual embeddings, together with a learnable aligning factor, has been critical to achieving optimal performance. 
Overall, the proposed TGRSIT system and Text-RSIR provide a practical multimodal solution for bandwidth-limited RS data delivery, enabling high-quality and semantically coherent HR reconstruction from compact transmitted representations.

\FloatBarrier
\bibliographystyle{cas-model2-names}
\bibliography{cas-refs}

\end{document}